\documentclass[prd,twocolumn,floatfix,amsmath,nofootinbib,amssymb,floatfix]{revtex4}
\usepackage{graphicx,color,dcolumn,booktabs,bm}
\usepackage{longtable,lscape}
\usepackage{pdfpages}
\usepackage{txfonts}
\usepackage{overpic}
\usepackage{amssymb}
\usepackage{makecell}
\usepackage{indentfirst}
\usepackage{feynmf}   %{feynmp}
\usepackage{slashed}  %for Feynman symbols
\usepackage{cases}
\usepackage{color}
\usepackage{multirow}
\usepackage{threeparttable}
\usepackage{epstopdf}
\usepackage{enumerate}
\usepackage{subfigure}
\usepackage{diagbox}
\usepackage{graphicx,color,dcolumn,booktabs,bm}
\usepackage{mathrsfs}
\usepackage{cancel}
\usepackage{float}
\usepackage[colorlinks,
            citecolor=blue,
            anchorcolor=red,
            menucolor=red,
            linkcolor=red,
            filecolor=red,
            runcolor=red,
            urlcolor=blue,
            frenchlinks=red]{hyperref}

\begin{document}

\title{Spin-dependent interactions and fine structure in the negative-parity singly heavy baryons }
\author{Zhen-Yu Li$^{1}$}
\email{zhenyvli@163.com }
\author{Guo-Liang Yu$^{2}$}
\email{yuguoliang2011@163.com }
\author{Zhi-Gang Wang$^{2}$ }
\email{zgwang@aliyun.com }
\author{Jian-Zhong Gu$^{3}$ }
\email{gujianzhong2000@aliyun.com }

\affiliation{$^1$ School of Physics and Electronic Science, Guizhou Education University, Guiyang 550018,
China\\$^2$ Department of Mathematics and Physics, North China Electric Power University, Baoding 071003,
China\\$^3$ China Institute of Atomic Energy, Beijing 102413,China}
\date{\today }

\begin{abstract}
In order to meet the high-precision measurement of the current baryon spectroscopy, for the first time, we rigorously analyze the spin-dependent interactions and the fine structure of the negative-parity singly heavy baryons in the relativized quark model, which was previously unfeasible in the three-quark system. This gains access to the exploration of the strong interactions dominated by the non-perturbative QCD, and reveals how the various forms of strong interactions in a baryon compete with each other, affect the evolution of the energy levels, cause the energy level splitting and contribute to the mixing effect responsible for the formation of the fine structures. It is shown that the rigorous calculation can nicely reproduce the data, the averaged deviation between the calculated and experimental energy levels is less than 5 MeV for the negative-parity singly heavy baryons. This confirms the reliability of the calculation and also helps make reasonable assignments for the observed negative-parity baryons. The large amount of data obtained by the rigorous calculations contains a wealth of interaction information and is helpful for both of the theoretical and experimental studies. The key to the rigorous calculation in this work is the proposal of a new method, namely the two-step Gaussian expansion method. This new method not only overcomes the long-standing unresolved problem in the relativized quark model, but also provides an effective approach for the high-precision calculation of other few-body systems such as the compact tetraquarks and pentaquarks, especially for the treatment of spin-orbit interactions and tensor interactions which actually appear ubiquitously in all of quantum many-body systems.
\end{abstract}

\maketitle

\section{Introduction}\label{sec1}
The heavy baryon spectroscopy has played an important role in the development of Quantum Chromodynamics (QCD)~\cite{F101}, especially in understanding the properties of strong interactions dominated by the non-perturbative QCD. In recent years, with the advances of the high energy facilities and detectors, some excited singly heavy baryons were discovered in groups~\cite{F201,F203,F204,F205,F206,F207,F208,F209,F210},
such as $\{\Omega_{c}(3000)^{0}$, $\Omega_{c}(3050)^{0}$, $\Omega_{c}(3065)^{0}$, $\Omega_{c}(3090)^{0}$, $\Omega_{c}(3120)^{0}\}$, $\{\Omega_{b}(6316)^{-}$, $\Omega_{b}(6330)^{-}$, $\Omega_{b}(6340)^{-}$, $\Omega_{b}(6350)^{-}\}$ and $\{\Xi_{c}(2923)^{+,0}$, $\Xi_{c}(2930)^{+,0}\}$. The mass values of these baryons in each group are very close and the mass gap is even less than 10 MeV, suggesting the existence of the fine structures in their excitation spectra.
Calculations showed that these baryons might belong to the $1P$-wave states~\cite{P02,P03,P05,P06,P07,P08,P09,P10,P11,P12,P01,P13,P14,P141,P15,P16,P17,P181,P18}. Exactly, they should be of the negative-parity. However, their spin values cannot be reliably assigned, and the close mass values in each group can not be reproduced precisely as well. How to explain these fine structures poses a great challenge for the theoretical analysis.

On the other hand, it is well known that the fine structure of the energy levels is generally caused by spin-dependent interactions, e.g., the fine structure in the atomic spectrum~\cite{S01} and the shell structure in Nuclei~\cite{S02}. Naturally, the fine structure in the heavy baryon spectra might also contain abundant information about the spin-dependent interactions. So, investigating the fine structure provides an opportunity to explore the details of these interactions within a heavy baryon.

For properly understanding the spin-dependent interactions and the fine structure in these heavy baryons, a reliable theoretical analysis is necessary, and it should be accurate enough. To this end, such an analysis must meet two requirements: (1) The model (or theory) should be precise enough, i.e., it should be QCD inspired and contain all possible and detailed spin-dependent interactions; (2) The model should be calculable with high-precision (or rigorously).
However, these two requirements present substantial challenges to current theoretical approaches, including lattice QCD~\cite{F101,P15,P20}, QCD sum rules~\cite{P21,P22,P221}, chiral perturbation theory~\cite{P23,P24}, Regge trajectories~\cite{P251,P25} and the quark potential models~\cite{P261,P26,P27,P28,P29,P31,P32,P33,P34,P35,P36,P37,P372}.

\begin{figure}[htbp]
\centering
\includegraphics[width=11cm,trim=3cm 11cm 2cm 4.6cm]{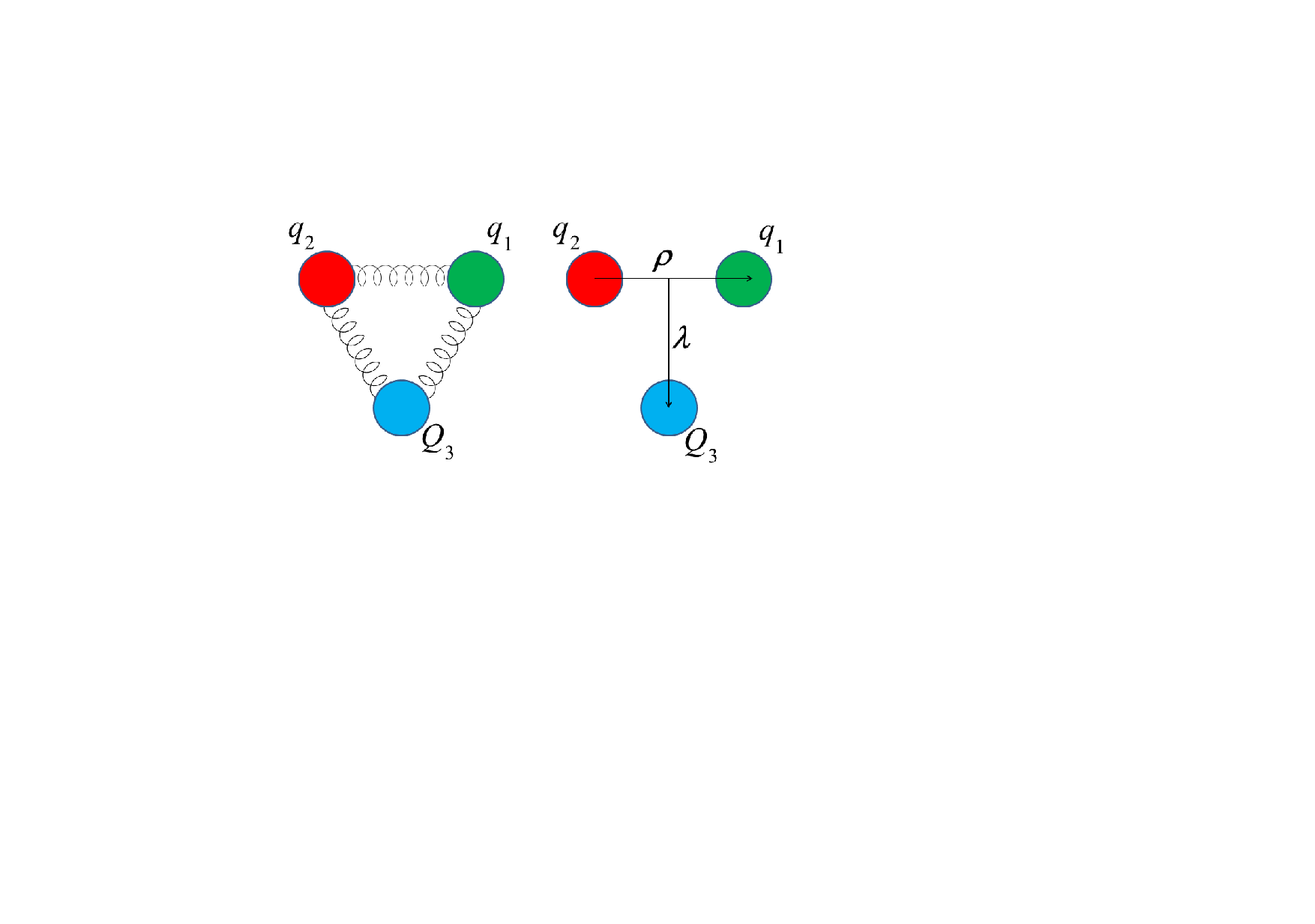}
\caption{The strong interaction within a baryon described by the relativized quark model is of the two-body interactions as illustrated on the left panel. While, the matrix element calculations of the interactions are performed in the specific Jacobi coordinates ($\boldsymbol\rho$, $\boldsymbol\lambda$) shown on the right panel. Here, $q_{1(2)}$ and $Q_{3}$ denote the light quark and the heavy quark, respectively. The difference between the used variables on the two panels causes a great difficulty in the real calculations. }
\label{fig1}
\end{figure}

As a representative quark potential model, the relativized quark model (RQM) might be recognized as a precise one since its Hamiltonian takes into account more physical factors concerning QCD and the relativistic correction than those in other quark models, which benefits the spin-dependent interactions there~\cite{P38,P39}. Namely, the RQM meets  requirement (1). Meeting requirement (2), however, the difficulties encounters far more serious than expected, for that the Hamiltonian in this model contains some rather complicated interaction terms.
For example, there are some spin-orbit terms in the RQM like $\langle\alpha|\frac{\textbf{s}_{i}\cdot(\mathbf{r}_{ij}\times\mathbf{p}_{i})}{2m^{2}_{i}r_{ij}}\frac{\partial\tilde{G}^{so(v)}_{ii}(r_{ij})}{\partial{r_{ij}}}|\beta\rangle$. Here, $|\alpha\rangle$ and $|\beta\rangle$ denote any two orbital excited states, which are defined based on the Jacobi coordinates ($\boldsymbol\rho$, $\boldsymbol\lambda$) as shown in Fig.~\ref{fig1}. While, $\mathbf{p}_{i}$ is defined in the Cartesian coordinates and $\mathbf{r}_{ij}$ $\equiv$ $\mathbf{r}_{i}$ - $\mathbf{r}_{j}$ represents the relative position vector between the $i$-th and $j$-th quarks ($\{i,j\}$ = $\{1,2\}$, $\{2,3\}$, or $\{3,1\}$), which is different from the Jacobi coordinates (see Fig.~\ref{fig1}). This difference makes the calculation extremely complicated, especially in cases of $\{i,j\}$ = $\{2,3\}$ and $\{3,1\}$.
By applying the coordinates transformation to $\mathbf{r}_{ij}$ and $\mathbf{p}_{i}$ further, the matrix element becomes
\begin{eqnarray}
\notag
&& \langle\alpha|\frac{\textbf{s}_{i}\cdot(\mathbf{r}_{ij}\times\mathbf{p}_{i})}{2m^{2}_{i}r_{ij}}\frac{\partial\tilde{G}^{so(v)}_{ii}(r_{ij})}{\partial{r_{ij}}}|\beta\rangle \\ \notag &&=\langle\alpha|\{\frac{\partial\tilde{G}^{so(v)}_{ii}(r_{ij})}{r_{ij}\partial{r_{ij}}}[\frac{A_{rij}A_{pi}}{2m^{2}_{i}}\textbf{\emph{l}}_{\rho}\cdot\mathbf{s}_{i}
 +\frac{B_{rij}B_{pi}}{2m^{2}_{i}}\textbf{\emph{l}}_{\lambda}\cdot\mathbf{s}_{i}\\
  &&+\frac{A_{rij}B_{pi}}{2m^{2}_{i}}(\boldsymbol\rho\times\mathbf{p}_{\lambda})\cdot\mathbf{s}_{i}
  +\frac{B_{rij}A_{pi}}{2m^{2}_{i}}(\boldsymbol\lambda\times\mathbf{p}_{\rho})\cdot\mathbf{s}_{i}]\}|\beta\rangle,
\label{eq1}
\end{eqnarray}
 with $\textbf{\emph{l}}_{\rho}\equiv\boldsymbol\rho\times\mathbf{p}_{\rho}$ and $\textbf{\emph{l}}_{\lambda}\equiv\boldsymbol\lambda\times\mathbf{p}_{\lambda}$. Here, $A_{rij}A_{pi}$, $B_{rij}B_{pi}$, $A_{rij}B_{pi}$ and $B_{rij}A_{pi}$ are the coefficients arising from the coordinate transformation.
 The form of Eq.~(\ref{eq1}) can conveniently describe the orbital excitation modes of baryons, which are dominant in low-energy regime.
 In Eq.~(\ref{eq1}), the latter two terms are proportional to $\boldsymbol\rho\times\mathbf{p}_{\lambda}$ and $\boldsymbol\lambda\times\mathbf{p}_{\rho}$, respectively. They are the so-called three-body spin-orbit potentials ($H^{so-3B}_{ij}$)~\cite{P40}. Then, the former two terms are named as the two-body interactions ($H^{so-2B}_{ij}$). These three-body terms were studied in the 1970s, but have not yet been fully and rigorously calculated so far~\cite{P39,P40,P41}. Actually, in addition to the spin-orbit terms, the tensor terms are also very difficult to calculate rigorously in the RQM, especially when the mixing effect is taken into account~\cite{P42,P43}. The rigorous calculation is a serious problem that the RQM encounters in the baryon system.

This problem has remained unsolved for nearly 50 years, leading to the RQM being unable to accurately analyze the baryon spectra and some fundamental physical issues having not been clarified either, such as the `spin-orbit puzzle'~\cite{P44}. For avoiding this problem, various approximations were put forward, such as the diquark approximation~\cite{P45}. But these approximations resulted in the unknown systematical errors and unreliable conclusions as discussed in Ref.~\cite{P46}. So, the approximate methods ultimately failed to meet the requirements of the high-precision analysis. Therefore, realizing the rigorous calculation of the RQM becomes the inevitable path for the current theoretical study.

In this work, we develop the Gaussian expansion method (GEM)~\cite{P47} to a two-step GEM for the first time. This two-step GEM enables us to work out these complicated matrix elements in calculation rigorously.
Not only can it overcome the long-standing major obstacle in the rigorous calculation of the baryon spectroscopy, but also lay the foundation for our accurate analysis of the spin-dependent interactions and the fine structure of the negative-parity heavy baryons in this work.

The remainder of this paper is organized as follows. In
Sec.~\ref{sec2}, the calculation scheme used in this work is introduced, including the wave functions, Jacobi coordinates, Hamiltonian of the RQM, GEM and calculation of the mixing effect. The results and discussion are presented in Sec.~\ref{sec3}. And Sec.~\ref{sec4} is reserved for the conclusions. The appendix in Sec.~\ref{sec5} presents the additional details on the calculation scheme and the complete calculation results. In particular, the two-step GEM is described there in detail.

\section{Calculation scheme}\label{sec2}
Commonly, the two light quarks in a singly heavy baryon are assumed to satisfy the flavor $SU(3)$ symmetry. As the requirement of the flavor $SU(3)$ subgroups, the singly heavy baryons belong to either a sextet ($\mathbf{6}_{F}$) of the flavor symmetric states ($\Sigma_{c}$, $\Xi'_{c}$, $\Omega_{c}$, $\Sigma_{b}$, $\Xi'_{b}$ and $\Omega_{b}$), or an anti-triplet ($\mathbf{\bar{3}}_{F}$) of the flavor antisymmetric states ($\Lambda_{c}$, $\Xi_{c}$, $\Lambda_{b}$ and $\Xi_{b}$). Calculations show that these fine structures concerned in this paper only exist in the $\mathbf{6}_{F}$ sector~\cite{P48}. We select the specific Jacobi coordinates $(\boldsymbol\rho$, $\boldsymbol\lambda)$ as shown in Fig.~\ref{fig1}, which is consistent with the flavor symmetry naturally. Then, the spin and orbit wave function is assumed to have the coupling scheme
\begin{eqnarray}
|(J^{P})_{j},L\rangle = |\{[(l_{\rho} l_{\lambda} )_{L}(s_{1}s_{2})_{s_{12}}]_{j} s_{3}\}_{J }\rangle,
\label{eq2}
\end{eqnarray}
with the parity $P=(-1)^{l_{\rho}+l_{\lambda}}$.
$l_{\rho}$($l_{\lambda}$), $L$ and $s_{12}$ are the quantum numbers of the relative orbital angular momentum $\textbf{\emph{l}}_{\rho}$ ($\textbf{\emph{l}}_{\lambda}$), total orbital angular momentum $\textbf{\emph{L}}$ and total spin of the light-quark pair $\mathbf{s}_{12}$, respectively. $j$ denotes the quantum number of the coupled angular momentum of $\textbf{\emph{L}}$ and $\textbf{s}_{12}$. Here, $(-1)^{l_{\rho}+s_{12}}=-1$ should be guaranteed for the $\mathbf{6}_{F}$ sector due to the total antisymmetry of the wave function of the two light quarks.
Simply, the orbital excited state is labeled with $(l_{\rho},l_{\lambda})L(J^{P})_{j}$.
The $(l_{\rho},l_{\lambda})L$ denotes the orbital excitation mode. It is commonly known as the $\lambda$-mode if $l_{\rho}=0$, or the $\rho$-mode if $l_{\lambda}=0$ in the literature~\cite{F101}.

In the RQM, the Hamiltonian is defined as~\cite{P38,P39},
\begin{eqnarray}\label{e1}
\notag
\hat{H} &&=\hat{H}_{0}+\hat{H}^{conf}+\hat{H}^{hyp}+\hat{H}^{so}\\
&&=\sum_{i=1}^{3}\sqrt{p_{i}^{2}+m_{i}^{2}}+\sum _{i<j}(\hat{H}^{conf}_{ij}+\hat{H}^{hyp}_{ij}+\hat{H}^{so}_{ij}).
\label{eq3}
\end{eqnarray}
The confinement term $\hat{H}^{conf}_{ij}$ includes a modified one-gluon-exchange potential and a smeared linear confinement potential. The hyperfine interaction $\hat{H}^{hyp}_{ij}$ consists of the tensor term $\hat{H}^{tens}_{ij}$ and the contact term $\hat{H}^{cont}_{ij}$. And the spin-orbit interaction $\hat{H}^{so}_{ij}$ can be divided into the color-magnetic term $\hat{H}^{so(v)}_{ij}$ and the Thomas-precession term $\hat{H}^{so(s)}_{ij}$. $\hat{H}^{tens}_{ij}$, $\hat{H}^{cont}_{ij}$, $\hat{H}^{so(v)}_{ij}$ and $\hat{H}^{so(s)}_{ij}$ belong to the spin-dependent interactions. As will be clear later, $\hat{H}^{mode}=\hat{H}_{0}+\hat{H}^{conf}$ determine the energy levels of the orbital excitation mode $(l_{\rho},l_{\lambda})L$. While, these spin-dependent terms contribute to the energy level splittings and the formation of the fine structures.

Based on some helpful studies~\cite{P01,P13,P14,P141} and our exploration in recent years in which the RQM was employed~\cite{P48,P49,P50,P51,P52,P53}, we summarize the following calculation scheme for the orbital excitation spectra:
(1) The calculation showed that for higher negative-parity states ($L>1$) such as the $F$-wave states, their excitation energies are rather high. While, the RQM might not accommodate such excited states with high energy since the RQM does not take into account high order excitations and explicit gluonic excitation in the Fock space and the RQM is in principle valid in low energy regime. Then, the negative-parity heavy baryon states are considered to originate only from the $P$-wave states.
(2) In theory, there are so many orbital excitation modes $(l_{\rho},l_{\lambda})L$ for the negative-parity $P$-wave states, such as $(1,0)P$, $(0,1)P$, $(1,2)P$, $(2,1)P$ and so on. The truly effective modes should be those with relatively lower excitation energies, which has been coined as the mechanism of the heavy-quark dominance (HQD)~\cite{P48,P51}. Based on this mechanism, the selected $P$-wave states consist of five $\lambda$-mode states $\{(0,1)P(\frac{1}{2}^{-})_{0}$,  $(0,1)P(\frac{1}{2}^{-})_{1}$, $(0,1)P(\frac{3}{2}^{-})_{1}$, $(0,1)P(\frac{3}{2}^{-})_{2}$, $(0,1)P(\frac{5}{2}^{-})_{2}\}$ and two $\rho$-mode states $\{(1,0)P(\frac{1}{2}^{-})_{1}$, $(1,0)P(\frac{3}{2}^{-})_{1}\}$ for the charmed baryons. For the bottom baryons, however, there are only five $\lambda$-mode states which are the same as those of the charmed baryons~\cite{P48,P52}.
(3) The above calculated orbital excited states do not correspond to the really physical observed states, because $L$ is not strictly a good quantum number. After taking into account the mixing effect, the physical observed states (or the baryon states) can be compared with these theoretical orbital excited states~\cite{P39,P40}.

The calculation process is as follows.
First, we calculate the Hamiltonian expectation value $\langle H\rangle \equiv\langle\alpha|\hat{H}|\alpha\rangle/\langle\alpha|\alpha\rangle$ in each orbital excited state $|\alpha\rangle\equiv|(l_{\rho},l_{\lambda})L(J^{P})_{j}\rangle$.
The calculations are performed in the generalized Gaussian base functions $|(n,\alpha)^{G}\rangle$ with $n=1-n_{max}$~\cite{P52}. In the non-orthogonal representation of $|(n,\alpha)^{G}\rangle$, the Hamiltonian $\hat{H}$ is represented as a matrix with $H_{nn'}$ = $\langle(n,\alpha)^{G}|\hat{H}|(n',\alpha)^{G}\rangle$. The solution belongs to a generalized matrix eigenvalue problem
\begin{eqnarray}
\begin{aligned}
\sum^{n_{max}}_{n'=1}(H_{nn'}-EN_{nn'})C_{n'}=0.
\label{eq4}
\end{aligned}
\end{eqnarray}
Here, $N_{nn'}$ comes from the non-orthogonality of the base functions. Then, the eigenvalue $E_{n\alpha}$ and the eigenvector $\{C_{n'}\}_{n\alpha}$ ($n'=1-n_{max}$) are obtained for the $|(n,\alpha)\rangle \equiv |(l_{\rho},l_{\lambda})nL(J^{P})_{j}\rangle$ state. Now, $n$ denotes the radial quantum number. As $n=1$ and $L=1$, the $E_{1\alpha}$ and  $\{C_{n'}\}_{1\alpha}$ correspond the Hamiltonian expectation value $\langle H\rangle$ and the wave function of the orbital excited state $(l_{\rho},l_{\lambda})1P(J^{P})_{j}$, respectively. These calculations have been finished rigorously in our previous work~\cite{P52}.
Some important issues of the calculation have also been extensively demonstrated, such as the improvement of the Hamiltonian parameters~\cite{P49}, the modification of the Gaussian size parameters~\cite{P50}, the convergence~\cite{P50} and the reliability of the calculation~\cite{P52}.

Next, we consider the mixing effect so as to obtain the physical observed states. The above orbital excited states with the same $J^{P}$ values form a subspace. Taking the charmed baryons as an example, there are three states with $J^{P}=\frac{1}{2}^{-}$, being noted as $|(0,1)1P(\frac{1}{2}^{-})_{0}\rangle\equiv|1,\frac{1}{2}^{-}\rangle$, $|(0,1)1P(\frac{1}{2}^{-})_{1}\rangle\equiv|2,\frac{1}{2}^{-}\rangle$ and $|(1,0)1P(\frac{1}{2}^{-})_{1}\rangle\equiv|3,\frac{1}{2}^{-}\rangle$. In this subspace, the diagonal elements of the Hamiltonian matrix, i.e., the expectation values $E_{1\alpha}$, can be obtained with the above method. For these off-diagonal matrix elements $\langle1,\frac{1}{2}^{-}|\hat{H}|2,\frac{1}{2}^{-}\rangle$, $\langle1,\frac{1}{2}^{-}|\hat{H}|3,\frac{1}{2}^{-}\rangle$ and $\langle2,\frac{1}{2}^{-}|\hat{H}|3,\frac{1}{2}^{-}\rangle$, they can be calculated in theory by using the $\{C_{n'}\}_{1\alpha}$ values obtained above. However, the real calculation is extremely difficult and encounters a historic challenge just as was discussed earlier.

In this work, we extend the GEM to the two-step GEM which enables us to rigorously work out these complicated matrix elements, especially the matrix elements of the tensor terms and the three-body spin-orbit terms.

The procedure for calculating some very complicated matrix elements like $H^{so-3B}_{ij}$ in Eq.~(\ref{eq1}) can be divided into two steps.
In the first step, the derivation of $(\boldsymbol\lambda\times\mathbf{p}_{\rho})\cdot\mathbf{s}_{i}|\beta\rangle$ or $(\boldsymbol\rho\times\mathbf{p}_{\lambda})\cdot\mathbf{s}_{i}|\beta\rangle$ is performed in the framework of the GEM. But, the result of this part contains non-standard Gaussian base functions. To work out the matrix elements, in the second step, we carry out the GEM specially for the non-standard Gaussian base functions and complete the calculation. That is the so-called two-step GEM.

Based on this key method, we obtained the complete Hamiltonian matrix through rigorous calculations. At last, the baryon states and their energy levels are obtained by diagonalizing the Hamiltonian matrix within the $J^{P}$ subspace. More details on the calculation scheme can be found in Sec.~\ref{sec5}.

\section{Results and discussion}\label{sec3}
By performing the calculation scheme, we get the abundant calculation data (see Sec.~\ref{sec5} Appendix G). Then, we have the chance to analyze the details of the strong interactions within the baryon system and the formation of the fine structures and give a reasonable interpretation of the experimental data.

\begin{figure}[htbp]
\centering
\includegraphics[width=8.5cm,trim=3cm 7cm 8cm 2cm]{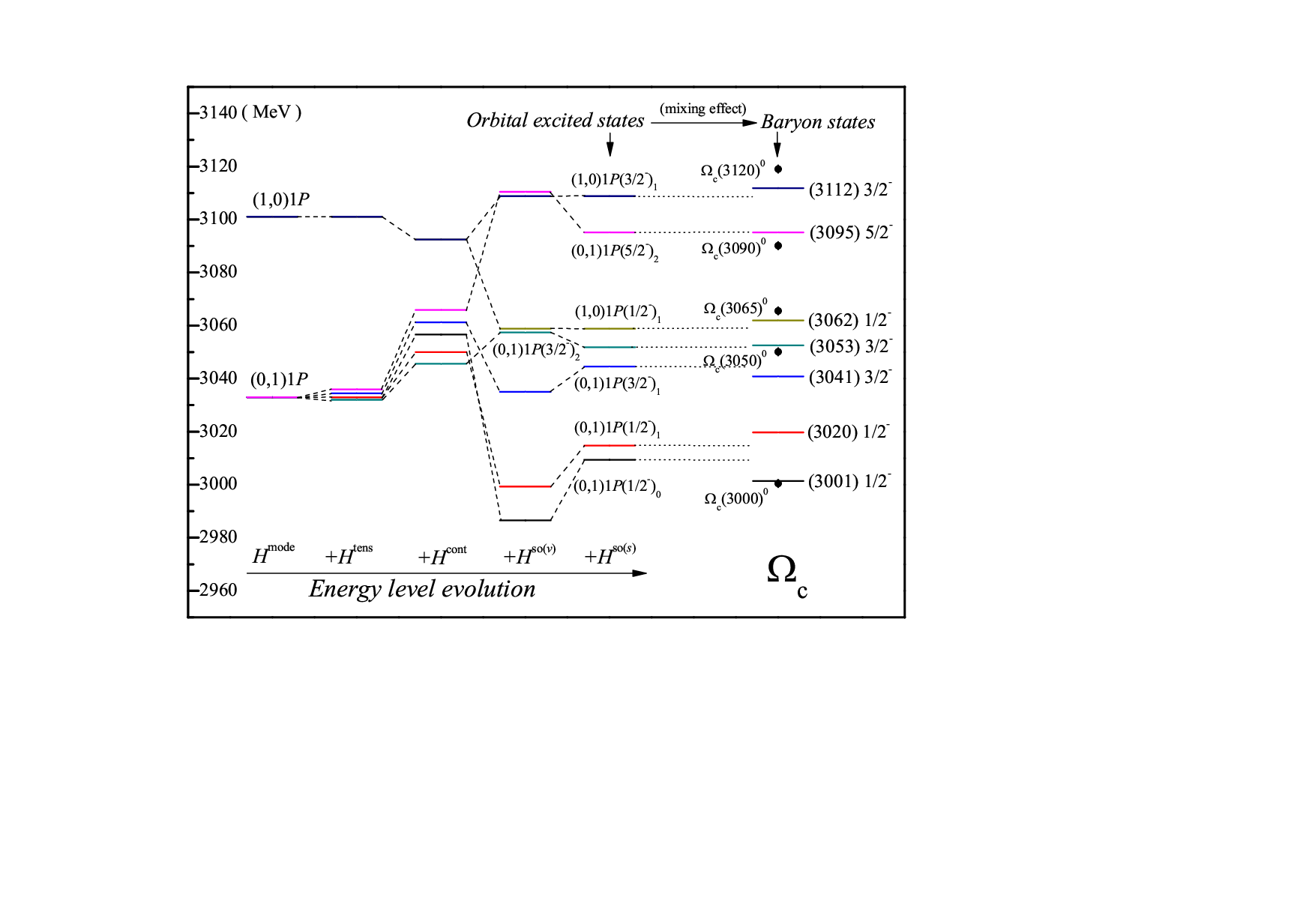}
\caption{The energy level evolution of the negative-parity $\Omega_{c}$ baryons. The contributions of the Hamiltonian terms to the energy levels are presented. $H^{mode}=H_{0}+H^{conf}$ determines the energy levels of the orbital excitation modes. The spin-dependent interactions cause the energy level splitting to different extents, respectively. This leads to the emergency of the rudimentary fine structure. By taking into account the mixing effect, the baryon states and the true fine structure are present. }
\label{fig2}
\end{figure}

(1) The contributions of the Hamiltonian terms to the energy levels.
Taking the $\Omega_{c}$ as an example (see Fig.~\ref{fig2}), by successively adding the $H^{mode}$, $H^{tens}$, $H^{cont}$, $H^{so(v)}$ and $H^{so(s)}$ terms, the evolution of the energy levels is presented for these $1P$-wave states. And the contributions of these spin-dependent terms to the energy levels are clearly displayed. The $H^{mode}$ term only determines the energy levels of the excitation modes. While, the spin-dependent terms cause the energy level splitting. It is shown that the effect of the tensor term $H^{tens}$ on the energy level splitting is very small, being only about 1 MeV. But, the contact term $H^{cont}$ significantly shifts these energy levels, and for the $(0,1)1P$ mode, the energy levels splitting occurs. With the addition of the color-magnetic term $H^{so(v)}$, the energy level splitting becomes very large for the $(0,1)1P$ and $(1,0)1P$ modes. While, the Thomas-precession term $H^{so(s)}$ counteracts the effect of the $H^{so(v)}$ term significantly for the $(0,1)1P$ mode and has a slight effect on the $(1,0)1P$ mode. The combined contribution of these spin-dependent terms makes the energy levels interleaved and shifted. This leads to the formation of the rudimentary fine structure, in which the spin-orbit terms play a more important role than the hyperfine terms.

(2) The contributions of the Hamiltonian terms to the mixing effect.
In fact, the rudimentary fine structure is not a true fine structure unless the mixing effect is taken into account. Calculations show that the spin-dependent terms lead to the configuration mixing. Taking the $\Omega_{c}$ baryons as an example, the contribution of each term to the off-diagonal matrix elements is presented in Table~\ref{tb1}, where the spin-orbit terms break up into the two-body interactions $H_{ij}^{so-2B}$ and the three-body ones $H_{ij}^{so-3B}$ as mentioned in Eq.~(\ref{eq1}). It is shown that the $H^{tens}$ and $H^{cont}$ (or $H^{so(v)}$ and $H^{so(s)}$) always partially cancel each other out.
As a whole, the three-body spin-orbit terms contribute more to the off-diagonal matrix elements than the other terms. Nevertheless, all of the spin-dependent interaction terms compete with and inhibit each other in a complex manner. All of them contribute to the mixing effect, none of them can be disregarded.

(3) The mixing effect.
In Fig.~\ref{fig2}, it is shown that the mixing effect causes a slight shift in the energy levels and leads to a nicely match between the predicted fine structure and the experimental data. So, the mixing effect is crucial for the formation of the fine structure.
\begin{table}[htbp]
\begin{ruledtabular}\caption{The off-diagonal matrix element (in MeV) of each Hamiltonian term for the negative-parity $\Omega_{c}$ baryons. In the $\frac{1}{2}^{-}$ subspace, the three states are abbreviated as $|(0,1)1P(\frac{1}{2}^{-})_{0}\rangle\equiv|1\rangle$, $|(0,1)1P(\frac{1}{2}^{-})_{1}\rangle\equiv|2\rangle$ and $|(1,0)1P(\frac{1}{2}^{-})_{1}\rangle\equiv|3\rangle$, respectively. And for the $\frac{3}{2}^{-}$ subspace, the three states are labeled as $|(0,1)1P(\frac{3}{2}^{-})_{1}\rangle\equiv|1'\rangle$, $|(0,1)1P(\frac{3}{2}^{-})_{2}\rangle\equiv|2'\rangle$ and $|(1,0)1P(\frac{3}{2}^{-})_{1}\rangle\equiv|3'\rangle$, respectively.}
\label{tb1}
\begin{tabular}{c c c c c c c c c c c c c c c c c c c c c}
\multirow{2}{*}{$\hat{O}$} & \multicolumn{3}{c}{$\Omega_{c}(\frac{1}{2}^{-})$} & \multicolumn{3}{c}{$\Omega_{c}(\frac{3}{2}^{-})$}  \\ \cline{2-4} \cline{5-7}
   &$\langle1|\hat{O}|2\rangle$  & $\langle1|\hat{O}|3\rangle$  & $\langle2|\hat{O}|3\rangle$ & $\langle1'|\hat{O}|2'\rangle$  & $\langle1'|\hat{O}|3'\rangle$  & $\langle2'|\hat{O}|3'\rangle$  \\ \hline
$\hat{H}_{12}^{tens}$ & 0  & 0  & 0   & 0  & 0  & 0\\
$\hat{H}_{23(31)}^{tens}$ &6.44& 6.19 & 3.72 & -4.97  & 2.97 & -6.94\\\hline
$\hat{H}_{12}^{cont}$ & 0  & 0  & 0   & 0  & 0  & 0\\
$\hat{H}_{23(31)}^{cont}$ & -5.53& -3.37& -4.74 &3.80  &-2.16  & 4.75  \\\hline
$\hat{H}_{12}^{so(v)-2B}$ & 0  & 0  & 0   & 0  & 0  & 0\\
$\hat{H}_{23(31)}^{so(v)-2B}$ & -11.48& 0& 5.40 &8.25& -5.18&0 \\\hline
$\hat{H}_{12}^{so(s)-2B}$ & 0  & 0  & 0   & 0  & 0  & 0\\
$\hat{H}_{23(31)}^{so(s)-2B}$ & 2.89 & 0 & -1.29 & -2.21 & 1.29 & 0\\\hline
$\hat{H}_{12}^{so(v)-3B}$ & 0 & 0 & 7.64 & 0 & -7.19 & 0  \\
$\hat{H}_{23(31)}^{so(v)-3B}$ & 3.99 & 0 & -15.93 & -2.84 & 15.03 & 0 \\\hline
$\hat{H}_{12}^{so(s)-3B}$ & 0& 0 & 11.22 & 0 &-10.99 & 0\\
$\hat{H}_{23(31)}^{so(s)-3B}$ & -0.29& 0 & 9.20 & 0.22 & -9.31 & 0 \\\hline
Total & -7.95 & 5.63 & 11.50 & 4.50 & -12.94 & -4.38  \\
\end{tabular}
\end{ruledtabular}
\end{table}

\begin{figure}[htbp]
\centering
\includegraphics[width=8.5cm,trim=2cm 7cm 7cm 2cm]{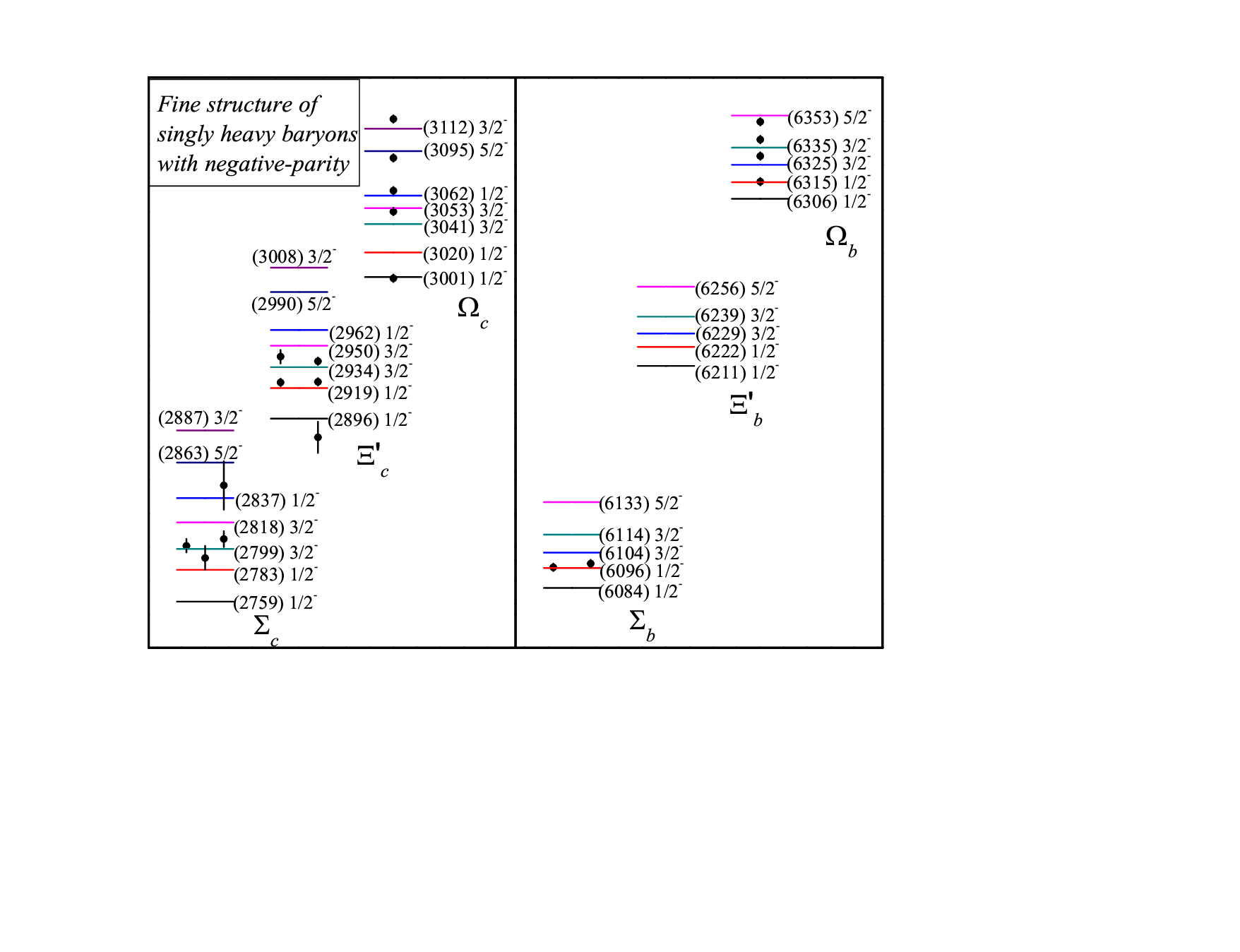}
\caption{Calculated baryon states of the negative-parity singly heavy baryons and their fine structures. The masses are measured in MeV. The solid black circles denote the observed baryon masses with experimental error bars. They in each family are $\{\Sigma_{c}(2800)^{++,+,0}$, $\Sigma_{c}(2846)^{0}\}$, $\{\Xi_{c}(2882)^{0}$, $\Xi_{c}(2923)^{+,0}$,  $\Xi_{c}(2930)^{+,0}\}$, $\{\Omega_{c}(3000)^{0}$, $\Omega_{c}(3050)^{0}$, $\Omega_{c}(3065)^{0}$, $\Omega_{c}(3090)^{0}$, $\Omega_{c}(3120)^{0}\}$, $\{\Sigma_{b}(6097)^{+,-}\}$, and $\{\Omega_{b}(6316)^{-}$, $\Omega_{b}(6330)^{-}$, $\Omega_{b}(6340)^{-}$, $\Omega_{b}(6350)^{-}\}$, respectively~\cite{F201,F210}.}
\label{fig3}
\end{figure}

(4) Fine structure and assignments of the observed baryons.
The fine structures of the negative-parity singly heavy baryons are presented in Fig.~\ref{fig3}. The masses of the observed baryons are also plotted in the figure. For the $\Omega_{c}$ and $\Omega_{b}$ baryons, their predicted fine structures match with the data nicely. The calculated masses of the other baryons are also close to the corresponding experimental energy levels. This indicates that the RQM can reproduce the fine structure nicely with the rigorous calculation. Another interesting feature is that the energy level gaps in the fine structure for the different families are approximately the same. In other words, these gaps hardly evolve along with the quark masses or the flavors.
Based on the nice match between the calculation and the data, we make the corresponding assignments for the 20 observed baryons which are plotted in Fig.~\ref{fig3} (also in the Table~\ref{tb19} of Appendix G). Additionally, the $\{\Xi_{b}(6227)^{0,-}\}$~\cite{F201} were once assigned to the $2S(\frac{1}{2}^{+})$ state of the $\mathbf{\bar{3}}_{F}$ sector in Ref.~\cite{P52}. Now, their measured masses are also close to that of the $(6229)\frac{3}{2}^{-}$ state. So, if they belong to the negative-parity baryons, they are the ideal candidates of the $(6229)\frac{3}{2}^{-}$ state of the $\mathbf{6}_{F}$ sector.

(5) Uncertainty and reliability.
For these negative-parity baryons, we evaluate the deviations of the calculated masses from their measured ones. The obtained arithmetic average deviation ($\sum_{i=1}^{N}|M_{cal.}-M_{exp.}|_{i}/N$) is less than 5 MeV (see Table~\ref{tb19}). While, the energy level gaps in the fine structure are generally about 10 MeV as mentioned in Sec.~\ref{sec1}. So, the calculation can distinguish the current experimental data, as shown in Fig.~\ref{fig3}. The overall systematic consistency between the prediction and the data confirms the reliability of the calculation scheme. By comparing the fine structures plotted in Fig.~\ref{fig3} with those in our previous work~\cite{P48,P51,P52}, one can conclude that the improvement of the calculation scheme and the rigorous calculation have effectively improved the prediction accuracy of the RQM. It also hints the RQM captures the important features of the strong interactions in the baryon system.

However, the uncertainty is a very hard issue to analyze. Actually, there are many potential sources of uncertainty that need to be clarified, such as the uncertainties coming from the model deficiency, model parameters, number of Gaussian basis functions considered, number of orbital basis states considered and channel-coupling effect.
They all result in the sizable deviation of the calculated results from the experimental data. Note that the model uncertainty caused by the model deficiency, especially the restriction of the Fock space and the rather schematic relativistic correction was analyzed~\cite{P38}. It was believed by the authors that the model uncertainty for heavy quark systems should be around 10 MeV. It is expected that for low-lying states of heavy quark systems the model uncertainty could be even smaller.
In this work, the improved calculation scheme and the achievement of the rigorous calculation have laid a reliable foundation for analyzing these potential factors in the future.

\section{Conclusions}\label{sec4}
The details of the strong interactions within baryons are not yet well understood in theory. An important reason lies in the lack of high-precision calculations with a precise model, which is actually a common difficulty that exists in the three-quark problem. In this work, to meet the high-precision measurement of the current baryon spectroscopy, the highly-precision and rigorous calculation with the RQM in a three-quark system has been achieved for the first time, which enables one to access the detailed analysis of the spin-dependent interactions and the fine structures in heavy baryons. The analysis of each Hamiltonian term reveals how the various forms of the strong interactions affect the evolution of the energy levels, the energy level splitting, the mixing effect and the formation of the fine structure. The high-precision calculations nicely reproduce the accurate experimental data, the averaged deviation between the calculated and experimental energy levels is less than 5 MeV for the negative-parity singly heavy baryons. Meanwhile, the spin quantum numbers of the observed baryons can be determined reliably. The nice match between the calculated and experimental data demonstrates the reliability of the RQM and the calculation scheme in this work, and illuminates the power and beauty of the QCD inspired RQM and the QCD at low energy.

In addition, there may be many potential sources of uncertainty that need to be clarified as discussed above. It is actually the ultimate question that any quark models need to answer, including the RQM. In this work, the improved calculation scheme of the RQM has been confirmed to be rational. Its reliability will be confirmed in more baryon families. This has laid a solid foundation for the discussion of the uncertainty in the future.

At last, for achieving the rigorous calculations, the two-step GEM has been developed for the first time in this work. As an innovation of the traditional methods, the two-step GEM not only overcomes the long-standing unresolved problem in the RQM, but also provides an effective approach for the study of other few-body systems such as the compact tetraquarks and pentaquarks, especially for the treatment of spin-orbit interactions and tensor interactions which actually appear ubiquitously in all of quantum many-body systems.

\begin{large}
\section*{Acknowledgements}
\end{large}

 This research was supported by the Natural Science Foundation of Guizhou Province-ZK[2024](General Project)650, the National Natural Science Foundation of China (Grant Nos. 11675265, 12575083), the Continuous Basic Scientific Research Project (Grant No. WDJC-2019-13) and the Leading Innovation Project (Grant No. LC 192209000701).

\section{Appendix}\label{sec5}

 This section is for the additional details on the Hamiltonian of the RQM, wave functions, Jacobi coordinates, HQD mechanism, GEM, off-digonal matrix elements, two-step GEM. The complete calculation results are presented, and Refs.~\cite{P38,P39,P40,P41,P47,P48,P49,P50,P51,P52} are also included.
\subsection{Hamiltonian of the relativized quark model}

In the relativized quark model, the Hamiltonian for a three-quark system is written as,
\begin{eqnarray}
\notag
H &&=H_{0}+\tilde{H}^{conf}+\tilde{H}^{hyp}+\tilde{H}^{so}\\
&&=\sum_{i=1}^{3}\sqrt{p_{i}^{2}+m_{i}^{2}}+\sum _{i<j}(\tilde{H}^{conf}_{ij}+\tilde{H}^{hyp}_{ij}+\tilde{H}^{so}_{ij}),
\end{eqnarray}
with
 \begin{eqnarray}
 \begin{aligned}
&\tilde{H}^{conf}_{ij}=G'_{ij}(r)+\tilde{S}_{ij}(r), \\
&\tilde{H}^{hyp}_{ij}=\tilde{H}^{tens}_{ij}+\tilde{H}^{cont}_{ij},\\
&\tilde{H}^{so}_{ij}=\tilde{H}^{so(v)}_{ij}+\tilde{H}^{so(s)}_{ij},
\end{aligned}
\end{eqnarray}
and
\begin{eqnarray}
\notag
\tilde{H}^{tens}_{ij} &&=-\frac{\textbf{s}_{i}\cdot\textbf{r}_{ij}\textbf{s}_{j}\cdot\textbf{r}_{ij}/r^{2}_{ij}}{m_{i}m_{j}}
\times(\frac{\partial^{2}}{\partial{r^{2}_{ij}}}-\frac{1}{r_{ij}}\frac{\partial}{\partial{r_{ij}}})\tilde{G}^{t}_{ij}\\
&&+\frac{\textbf{s}_{i}\cdot\textbf{s}_{j}}{3m_{i}m_{j}}
\times(\frac{\partial^{2}}{\partial{r^{2}_{ij}}}-\frac{1}{r_{ij}}\frac{\partial}{\partial{r_{ij}}})\tilde{G}^{t}_{ij},
\end{eqnarray}
\begin{eqnarray}
\tilde{H}^{cont}_{ij}=\frac{2\textbf{s}_{i}\cdot\textbf{s}_{j}}{3m_{i}m_{j}}\nabla^{2}\tilde{G}^{c}_{ij},
\end{eqnarray}
\begin{eqnarray}
\notag
\tilde{H}^{so(v)}_{ij}&&=\frac{\textbf{s}_{i}\cdot(\mathbf{r}_{ij}\times\mathbf{p}_{i})}{2m^{2}_{i}r_{ij}}\frac{\partial\tilde{G}^{so(v)}_{ii}}{\partial{r_{ij}}}+
\frac{\textbf{s}_{j}\cdot(-\mathbf{r}_{ij}\times\mathbf{p}_{j})}{2m^{2}_{j}r_{ij}}\frac{\partial\tilde{G}^{so(v)}_{jj}}{\partial{r_{ij}}}\\
&&+\frac{\textbf{s}_{i}\cdot(-\mathbf{r}_{ij}\times\mathbf{p}_{j})+\textbf{s}_{j}\cdot(\mathbf{r}_{ij}\times\mathbf{p}_{i})}{m_{i}m_{j}r_{ij}}\frac{\partial\tilde{G}^{so(v)}_{ij}}{\partial{r_{ij}}},
\end{eqnarray}
\begin{eqnarray}
\tilde{H}^{so(s)}_{ij}=-\frac{\textbf{s}_{i}\cdot(\mathbf{r}_{ij}\times\mathbf{p}_{i})}{2m^{2}_{i}r_{ij}}\frac{\partial\tilde{S}^{so(s)}_{ii}}{\partial{r_{ij}}}-
\frac{\textbf{s}_{j}\cdot(-\mathbf{r}_{ij}\times\mathbf{p}_{j})}{2m^{2}_{j}r_{ij}}\frac{\partial\tilde{S}^{so(s)}_{jj}}{\partial{r_{ij}}}.
\end{eqnarray}
In the formulas above, $G'_{ij}$, $\tilde{G}^{t}_{ij}$, $\tilde{G}^{c}_{ij}$, $\tilde{G}^{so(v)}_{ij}$ and $\tilde{S}^{so(s)}_{ii}$ have been modified with the momentum-dependent factors as follows,
\begin{eqnarray}
\begin{aligned}
&G'_{ij}=(1+\frac{p^{2}_{ij}}{E_{i}E_{j}})^{\frac{1}{2}}\tilde{G}_{ij}(r_{ij})(1+\frac{p^{2}_{ij}}{E_{i}E_{j}})^{\frac{1}{2}}, \\
&\tilde{G}^{t}_{ij}=(\frac{m_{i}m_{j}}{E_{i}E_{j}})^{\frac{1}{2}+\epsilon_{t}}\tilde{G}_{ij}(r_{ij})(\frac{m_{i}m_{j}}{E_{i}E_{j}})^{\frac{1}{2}+\epsilon_{t}},\\
&\tilde{G}^{c}_{ij}=(\frac{m_{i}m_{j}}{E_{i}E_{j}})^{\frac{1}{2}+\epsilon_{c}}\tilde{G}_{ij}(r_{ij})(\frac{m_{i}m_{j}}{E_{i}E_{j}})^{\frac{1}{2}+\epsilon_{c}},\\
&\tilde{G}^{so(v)}_{ij}=(\frac{m_{i}m_{j}}{E_{i}E_{j}})^{\frac{1}{2}+\epsilon_{so(v)}}\tilde{G}_{ij}(r_{ij})(\frac{m_{i}m_{j}}{E_{i}E_{j}})^{\frac{1}{2}+\epsilon_{so(v)}}, \\
&\tilde{S}^{so(s)}_{ii}=(\frac{m_{i}m_{i}}{E_{i}E_{i}})^{\frac{1}{2}+\epsilon_{so(s)}}\tilde{S}_{ij}(r_{ij})(\frac{m_{i}m_{i}}{E_{i}E_{i}})^{\frac{1}{2}+\epsilon_{so(s)}},
\end{aligned}
\end{eqnarray}
where $E_{i}=\sqrt{m^{2}_{i}+p^{2}_{ij}}$ is the relativistic kinetic energy, and $p_{ij}$ is the momentum magnitude of either of the
quarks in the center-of-mass frame of the $ij$-th quark subsystem.

$\tilde{G}_{ij}(r_{ij})$ and $\tilde{S}_{ij}(r_{ij})$ are obtained by the smearing transformations of the one-gluon exchange potential $G(r)=-\frac{4\alpha_{s}(r)}{3r}$ and
linear confinement potential $S(r)=\tilde{b}r+\tilde{c}$, respectively,
\begin{eqnarray}
&\tilde{G}_{ij}(r_{ij})=\textbf{F}_{i}\cdot\textbf{F}_{j} \sum^{3}_{k=1}\frac{2\alpha_{k}}{\sqrt{\pi}r_{ij}}\int^{\tau_{kij}r_{ij}}_{0}e^{-x^{2}}\mathrm{d}x,
\end{eqnarray}
\begin{eqnarray}
\notag
\tilde{S}_{ij}(r_{ij}) &&=-\frac{3}{4}\textbf{F}_{i}\cdot\textbf{F}_{j} \{\tilde{b}r_{ij}[\frac{e^{-\sigma^{2}_{ij}r^{2}_{ij}}}{\sqrt{\pi}\sigma_{ij} r_{ij}}\\
&&+(1+\frac{1}{2\sigma^{2}_{ij}r^{2}_{ij}})\frac{2}{\sqrt{\pi}}\int^{\sigma_{ij}r_{ij}}_{0}e^{-x^{2}}\mathrm{d}x]+\tilde{c} \},
\end{eqnarray}
with
\begin{eqnarray}
\begin{aligned}
&\tau_{kij}=\frac{1}{\sqrt{\frac{1}{\sigma^{2}_{ij}}+\frac{1}{\gamma^{2}_{k}}}}, \\
&\sigma_{ij}=\sqrt{s^{2}_{0}(\frac{2m_{i}m_{j}}{m_{i}+m_{j}})^{2}+\sigma^{2}_{0}[\frac{1}{2}(\frac{4m_{i}m_{j}}{(m_{i}+m_{j})^{2}})^{4}+\frac{1}{2}]}.
\end{aligned}
\end{eqnarray}
Here $\alpha_{k}$ and $\gamma_{k}$ are constants. $\textbf{F}_{i}\cdot\textbf{F}_{j}$ stands for the inner product of the color matrices of quarks $i$ and $j$. For the baryon, $\langle\textbf{F}_{i}\cdot\textbf{F}_{j}\rangle=-2/3$. All of the parameters in these formulas are completely consistent with those in our previous works~\cite{P49,P50}. Their values are listed in Table~\ref{tta001}.

\begin{table*}[htbp]
\begin{ruledtabular}\caption{Parameters of the relativized quark model in this work. Their values are the same as those in Ref.~\cite{P38}, apart from $\tilde{b}$ and $\tilde{c}$~\cite{P49}. }
\label{tta001}
\begin{tabular}{c c c c c c c c c c c}
$m_{u}$/$m_{d}$(GeV)& $m_{s}$(GeV)   & $m_{c}$(GeV) & $m_{b}$(GeV) & $\gamma_{1}$(GeV) & $\gamma_{2}$(GeV) & $\gamma_{3}$(GeV)  & $\tilde{b}$(GeV$^{2}$) & $\tilde{c}$(GeV) \\
0.22 & 0.419  & 1.628 & 4.977& $1/2$ & $\sqrt{10}/2$ & $\sqrt{1000}/2$  & 0.14 & -0.198   \\\hline
$\epsilon_{c}$ & $\epsilon_{t}$   & $\epsilon_{SO(v)}$ & $\epsilon_{SO(s)}$  & $\alpha_{1}$ & $\alpha_{2}$ & $\alpha_{3}$  & $\sigma_{0}$(GeV)& $\tilde{s}$   \\
-0.168 & 0.025  & -0.035 & 0.055  & 0.25 & 0.15 & 0.20 & 1.8 & 1.55    \\
\end{tabular}
\end{ruledtabular}
\end{table*}

\subsection{Wave function}
For a singly heavy baryon system, the heavy-quark is decoupled from the two light-quarks in the heavy quark limit. With the requirement of the flavor $SU(3)$ subgroups for the light-quark pair, the singly heavy baryons belong to either a sextet ($\mathbf{6}_{F}$) of the flavor symmetric states,
 \begin{eqnarray}
\begin{aligned}
&\Sigma_{Q}=(uu)Q,~\frac{1}{\sqrt{2}}(ud+du)Q,~(dd)Q, \\
&\Xi_{Q}^{'}=\frac{1}{\sqrt{2}}(us+su)Q,~\frac{1}{\sqrt{2}}(ds+sd)Q,\\
&\Omega_{Q}=(ss)Q,
\end{aligned}
\end{eqnarray}
or an anti-triplet ($\mathbf{\bar{3}}_{F}$) of the flavor antisymmetric states,
\begin{eqnarray}
\begin{aligned}
&\Lambda_{Q}=~\frac{1}{\sqrt{2}}(ud-du)Q, \\
&\Xi_{Q}=\frac{1}{\sqrt{2}}(us-su)Q,~\frac{1}{\sqrt{2}}(ds-sd)Q.
\end{aligned}
\end{eqnarray}
Here $u$, $d$ and $s$ denote up, down and strange quark, respectively. $Q$ denotes charm ($c$) quark or bottom ($b$) quark.

\begin{figure}[htbp]
\centering
\includegraphics[width=8.5cm]{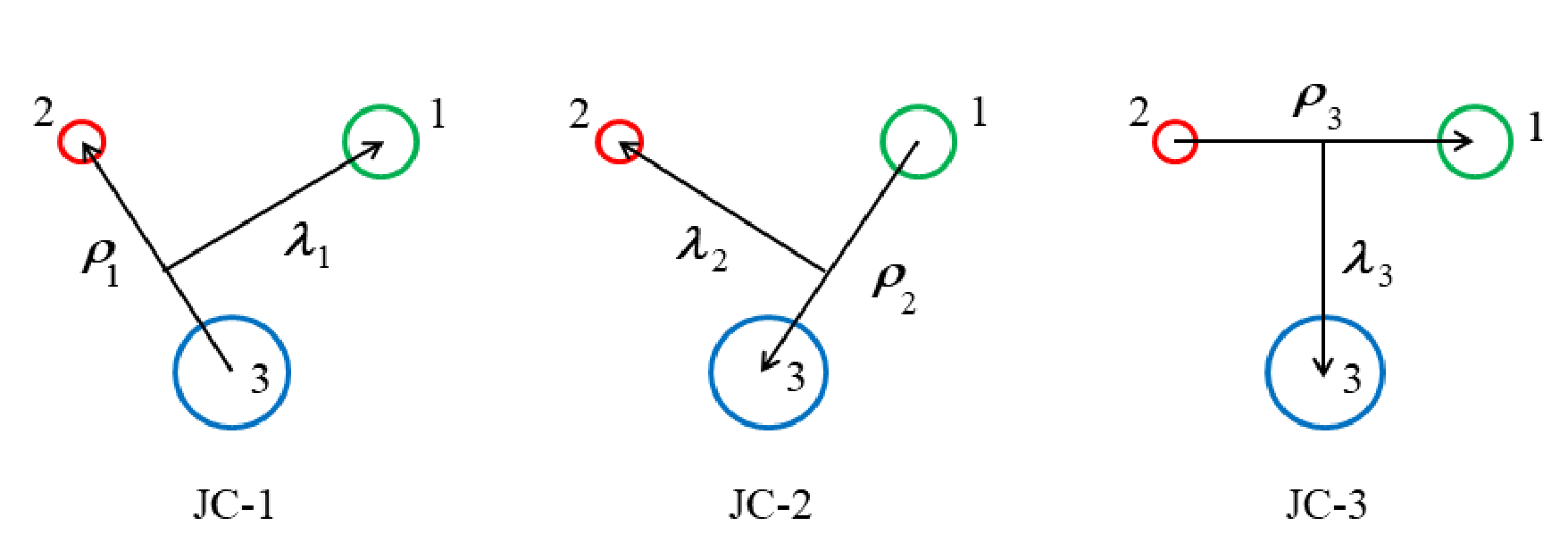}
\caption{There are 3 channels of the Jacobi coordinates for a three-quark system, labeled with ($\boldsymbol\rho_{k}$, $\boldsymbol\lambda_{k}$) ($k$=1, 2, 3). The channel 3 (JC-3) is selected for defining the orbital excited states of a singly heavy baryon. The quarks are numbered for ease of use in calculations, and the 3rd quark refers specifically to the heavy quark. In this work, all the calculations are based on the JC-3.}
\label{fig4}
\end{figure}

For describing the internal orbital motion of the singly heavy baryon, we select the specific Jacobi coordinates (JC-3) as shown in Fig.~\ref{fig4}, which is consistent with the above reservation about the flavor wave function naturally.
In this work, the Jacobi coordinates are defined as
\begin{eqnarray}
\begin{aligned}
&\boldsymbol\rho_{i}=\textbf{r}_{jk}=\textbf{r}_{j}-\textbf{r}_{k}, \\
&\boldsymbol\lambda_{i}=\textbf{r}_{i}-\frac{m_{j}\textbf{r}_{j}+m_{k}\textbf{r}_{k}}{m_{j}+m_{k}},\\
&\boldsymbol R_{i}=\frac{m_{i}\textbf{r}_{i}+m_{j}\textbf{r}_{j}+m_{k}\textbf{r}_{k}}{m_{i}+m_{j}+m_{k}}\equiv \mathbf{0},
\label{e13}
\end{aligned}
\end{eqnarray}
where $\{i$, $j$, $k\}$ = $\{$1, 2, 3$\}$, $\{$2, 3, 1$\}$ or $\{$3, 1, 2$\}$. $\textbf{r}_{i}$ and $m_{i}$ denote the position vector and the mass of the $i$th quark, respectively. $\textbf{\emph{R}}_{i}\equiv \textbf{0}$ means that the kinetic energy of the center of mass is not considered. Specially, for the JC-3 in Fig.~\ref{fig4}, the following convention is used in this work, $\boldsymbol\rho_{3}\equiv \boldsymbol\rho$ and $\boldsymbol\lambda_{3}\equiv \boldsymbol\lambda$.

Based on the JC-3, the spin and orbit wave function is assumed to have the coupling scheme
\begin{eqnarray}
|(J^{P})_{j},L\rangle = |\{[(l_{\rho} l_{\lambda} )_{L}(s_{1}s_{2})_{s_{12}}]_{j} s_{3}\}_{J }\rangle,
\end{eqnarray}
which is the same as Eq.~(\ref{eq2}).

\subsection{Orbital excitation and Heavy-quark dominance}
For the $L$-wave excitation with $\textbf{\emph{L}}$=$\textbf{\emph{l}}_{\rho}$+$\textbf{\emph{l}}_{\lambda}$, there are an infinite number of orbital excitation modes.
Taking $L=1$ as an example, the excitation modes $(l_{\rho},l_{\lambda})L$ are $(1,0)P$, $(0,1)P$, $(1,1)P$, $(2,1)P$, $(1,2)P$, $(2,2)P$, and so on.
The calculation cannot cover all the states. Moreover, the RQM does not take into account high order excitations and explicit gluonic excitation in the Fock space and the RQM is in principle valid in low energy regime.
We assume that the excitation mode with the lowest energy levels is the most stable and has the greatest probability of being observed experimentally, which dominates the structure of the excitation spectrum. Based on this assumption, we investigated the orbital excitation modes of singly and doubly heavy baryons.
It turns out that the excitation mode with lower energy levels is always associated with the heavy quark(s), and the splitting of the energy levels is suppressed by the heavy quark(s) as well. In other words, the heavy quarks dominate the orbital excitation of singly and doubly heavy baryons, and determine the structures of their excitation spectra.
For the bottom baryons, the lower excitation modes come mainly from the $\lambda$-modes, which is associated with the heavy quark. But there are exceptions for the $1P$-wave orbital excitation of $\Sigma_{c}$, $\Xi_{c}^{'}$ and $\Omega_{c}$ baryons, where the $\rho$-modes intrude the energy region of the $\lambda$-modes. In this case, the HQD mechanism is slightly broken, since $c$ quark is not heavy enough. But, the HQD mechanism is generally effective. We coined it as the mechanism of heavy-quark dominance\cite{P48,P51}.

Note that in 1978 Isgur and Karl made a discussion about the symmetry breaking in baryons [Phys. Lett. B \textbf{74} 4, 4-5 (1978)], where a similar conclusion was obtained for the strange baryons. The HQD mechanism and its breaking could be viewed as an extension of that issue to the heavy baryons.

\subsection{GEM and Hamiltonian expectation values}
Given a set of the orbital quantum numbers $\{l$, $m\}$, the Gaussian basis function $|(nlm)^{G} \rangle$ is commonly written in  position space as
 \begin{eqnarray}
\begin{aligned}
 &\phi^{G}_{nlm}(\textbf{r})=\phi^{G}_{nl}(r)Y_{lm}(\hat{\textbf{r}}),\\
 &\phi^{G}_{nl}(r)=N_{nl}r^{l}e^{-\nu_{n}r^{2}},\\
 &N_{nl}=\sqrt{\frac{2^{l+2}(2\nu_{n})^{l+3/2}}{\sqrt{\pi}(2l+1)!!}},
\end{aligned}
\end{eqnarray}
with
\begin{eqnarray}
\begin{aligned}
& \nu_{n}=\frac{1}{r^{2}_{n}},\\
& r_{n}=r_{1}a^{n-1}\ \ \ (n=1,\ 2,\ ...,\ n_{max}).
\end{aligned}
\end{eqnarray}
$\{r_{1}, a, n_{max}\}$ (or equivalently $\{n_{max},r_{1},r_{n_{max}}\}$) are the Gaussian size
parameters and commonly related to the scale in question.  The optimized values of $\{n_{max}=10$, $r_{1}=0.18$ GeV$^{-1}$, $r_{n_{max}}=15$ GeV$^{-1}\}$ are finally selected for the heavy baryons in this work. Details can be found in Refs.~\cite{P49,P50}.

The $\{\phi^{G}_{nlm}\}$ forms a set of finite-dimensional, non-orthogonal and complete bases,
 \begin{eqnarray}
\begin{aligned}
 &N_{n,n'}=\langle\phi^{G}_{nlm}|\phi^{G}_{n'lm}\rangle=(\frac{2\sqrt{\nu_{n}\nu_{n'}}}{\nu_{n}+\nu_{n'}})^{l+\frac{3}{2}},\\
 &1=\sum^{n_{max}}_{n=1}\sum^{n_{max}}_{n'=1}|\phi^{G}_{nlm}\rangle(N^{-1})_{nn'}\langle\phi^{G}_{n'lm}|.
 \end{aligned}
\end{eqnarray}
An arbitrary wave function $\psi_{lm}(\mathbf{r})$ can be expanded in a set of definite orbital quantum states,
 \begin{eqnarray}
\begin{aligned}
 &|\psi_{lm}\rangle=\sum^{n_{max}}_{n,n'=1}|\phi^{G}_{nlm}\rangle(N^{-1})_{nn'}\langle\phi^{G}_{n'lm}|\psi_{lm}\rangle
 \equiv\sum^{n_{max}}_{n=1}C_{n}|\phi^{G}_{nlm}\rangle.
 \end{aligned}
\end{eqnarray}

For a baryon system, based on the Jacobi coordinates, its spin-orbital state is defined as $|\{[(l_{\rho} l_{\lambda} )_{L}(s_{1}s_{2})_{s_{12}}]_{j} s_{3}\}_{JM_{J} }\rangle\equiv|\alpha\rangle_{3}$ (corresponding to the JC-3). Then, the corresponding generalized Gaussian basis function has the form below,
\begin{eqnarray}
\notag
|(\tilde{n},\alpha)^{G}_{3}\rangle &&=\sum_{\{m_{\xi}\}}\{CG_{\xi}\}\times|(n_{\rho}l_{\rho}m_{\rho})^{G} \rangle\otimes|(n_{\lambda}l_{\lambda}m_{\lambda})^{G} \rangle\\
&&\otimes|s_{1}m_{s_{1}}\rangle\otimes|s_{2}m_{s_{2}}\rangle\otimes|s_{3}m_{s_{3}}\rangle,
\label{eq23}
\end{eqnarray}
where $\{m_{\xi}\}$ denote all the 3rd components of the orbital angular momenta and spins, $\{CG_{\xi}\}$ are the products of all the C-G coefficients. $\tilde{n}$ is obtained by combining $n_{\rho}$ and $n_{\lambda}$, e.g., $\tilde{n}=(n_{\rho}-1)\times n_{max}+n_{\lambda}$ as $n_{\rho(\lambda)}=1,\cdot\cdot\cdot,n_{max}$.

The non-orthogonal and complete relations are as follows,
\begin{eqnarray}
\begin{aligned}
 &N_{\tilde{n},\tilde{n}'}=\langle (\tilde{n},\alpha)^{G}_{3}|(\tilde{n}',\alpha)^{G}_{3}\rangle\\
 &=(\frac{2\sqrt{\nu_{n_{\rho}}\nu_{n_{\rho}'}}}{\nu_{n_{\rho}}+\nu_{n_{\rho}'}})^{l_{\rho}+\frac{3}{2}}
 \times (\frac{2\sqrt{\nu_{n_{\lambda}}\nu_{n_{\lambda}'}}}{\nu_{n_{\lambda}}+\nu_{n_{\lambda}'}})^{l_{\lambda}+\frac{3}{2}},\\
 &1=\sum^{n^{2}_{max}}_{\tilde{n}=1}\sum^{n^{2}_{max}}_{\tilde{n}'=1}|(\tilde{n},\alpha)^{G}_{3}\rangle(N^{-1})_{\tilde{n}\tilde{n}'}\langle (\tilde{n}',\alpha)^{G}_{3}|.
 \end{aligned}
\end{eqnarray}

Thus, the evaluation of the expectation value $\langle\alpha|\hat{H}|\alpha\rangle_{3}/\langle\alpha|\alpha\rangle_{3}$ can be performed in the generalized Gaussian basis function $|(\tilde{n},\alpha)^{G}_{3}\rangle$, and the solution belongs to a generalized matrix eigenvalue problem
\begin{eqnarray}
\begin{aligned}
\sum^{n^{2}_{max}}_{\tilde{n}'=1}(H_{\tilde{n}\tilde{n}'}-EN_{\tilde{n}\tilde{n}'})C_{\tilde{n}'}=0.
\label{eq25}
\end{aligned}
\end{eqnarray}
The matrix element of an operator $\hat{H}$ reads,
\begin{eqnarray}
\notag
H_{\tilde{n}\tilde{n}'} &&=\langle(\tilde{n},\alpha)^{G}_{3}|\hat{H}|(\tilde{n}',\alpha)^{G}_{3}\rangle\\
\notag
&&=\sum_{\{m_{\xi}\},\{m_{\xi}'\}}\{CG_{\xi}\}\times\{CG_{\xi'}\}\times\langle(n_{\rho}l_{\rho}m_{\rho})^{G}|\langle(n_{\lambda}l_{\lambda}m_{\lambda})^{G}|\\
\notag
&& \times\langle s_{1}m_{s_{1}}|\langle s_{2}m_{s_{2}}|\langle s_{3}m_{s_{3}}|\hat{H}|s_{1}m_{s_{1}}'\rangle|s_{2}m_{s_{2}}'\rangle|s_{3}m_{s_{3}}'\rangle \\
\notag
&&\times|(n_{\rho}'l_{\rho}m_{\rho}')^{G} \rangle|(n_{\lambda}'l_{\lambda}m_{\lambda}')^{G} \rangle\\
&&\equiv\sum_{\{m_{\xi},m_{\xi}'\}}\{CG_{\xi}\times CG_{\xi'}\}\times H_{(n_{\rho}n_{\lambda},n_{\rho}'n_{\lambda}');(m_{s_{1,2,3}},m_{s_{1,2,3}}')}.
\label{eq26}
\end{eqnarray}
The rigorous calculation of $H_{(n_{\rho}n_{\lambda},n_{\rho}'n_{\lambda}');(m_{s_{1,2,3}},m_{s_{1,2,3}}')}$ has been finished in our previous work~\cite{P52}.

\subsection{Off-diagonal matrix elements of the Hamiltonian}
There are several different orbital excited states with the same $J^{P}$ values, which form a subspace and are denoted in order as $|1,J^{P}\rangle$, $|2,J^{P}\rangle$ and so on. The Hamiltonian $\hat{H}$ is represented as a matrix in this space with the element $\langle i,J^{P}|\hat{H}|j,J^{P}\rangle$ (or $\langle \alpha|\hat{H}|\beta\rangle/\sqrt{\langle\alpha|\alpha\rangle\langle\beta|\beta\rangle}$). The diagonal element with $\alpha=\beta$ is the expected value $\langle\alpha|\hat{H}|\alpha\rangle_{3}/\langle\alpha|\alpha\rangle_{3}$ as obtained above. The off-diagonal elements with $\alpha\neq \beta$ are obtained in the basis of the Gaussian base functions as follow,
\begin{eqnarray}
\notag
\frac{\langle \alpha|\hat{H}|\beta\rangle}{\sqrt{\langle\alpha|\alpha\rangle\langle\beta|\beta\rangle}} &&=\frac{\sum C^{*}_{\tilde{n},\alpha}\times\langle(\tilde{n},\alpha)^{G}_{3}|\hat{H}|(\tilde{n}',\beta)^{G}_{3}\rangle\times C_{\tilde{n}',\beta}}{\sqrt{\langle\alpha|\alpha\rangle\langle\beta|\beta\rangle}}\\
&&=\frac{\sum C^{*}_{\tilde{n},\alpha}\times H_{(\tilde{n},\alpha)(\tilde{n}',\beta)}\times C_{\tilde{n}',\beta}}{\sqrt{\langle\alpha|\alpha\rangle\langle\beta|\beta\rangle}},
\end{eqnarray}
with
\begin{eqnarray}
\langle\alpha|\alpha\rangle =\sum C^{*}_{\tilde{n},\alpha}\times N_{(\tilde{n},\alpha)(\tilde{n}',\alpha)}\times C_{\tilde{n}',\alpha}.
\end{eqnarray}

Here, $\sum$ is an abbreviation for $\sum_{\tilde{n},\tilde{n}'}$. The eigenvectors $\{C_{\tilde{n},\alpha}\}$ and $\{C_{\tilde{n},\beta}\}$ can be obtained from the solution of the generalized matrix eigenvalue problem in Eq.~(\ref{eq25}). Then, $H_{(\tilde{n},\alpha)(\tilde{n}',\beta)}$ can be worked out for most of the Hamiltonian terms, apart from the tensor terms and the three-body spin-orbit terms.

\subsection{Two-step GEM for three-body spin-orbit terms}
The off-diagonal matrix elements of the three-body spin-orbit terms and the tensor terms have not been worked out through rigorous calculation for a long time. In this work, we have developed the GEM to the two-step GEM, and finally solved this problem.

In the spin-orbit terms, the color-magnetic term $H^{so(v)}_{ij}$ reads,
\begin{eqnarray}\label{e8}
\notag
\tilde{H}^{so(v)}_{ij}&&=\frac{\textbf{s}_{i}\cdot(\mathbf{r}_{ij}\times\mathbf{p}_{i})}{2m^{2}_{i}r_{ij}}\frac{\partial\tilde{G}^{so(v)}_{ii}}{\partial{r_{ij}}}+
\frac{\textbf{s}_{j}\cdot(-\mathbf{r}_{ij}\times\mathbf{p}_{j})}{2m^{2}_{j}r_{ij}}\frac{\partial\tilde{G}^{so(v)}_{jj}}{\partial{r_{ij}}}\\
\notag
&&+\frac{\textbf{s}_{i}\cdot(-\mathbf{r}_{ij}\times\mathbf{p}_{j})
+\textbf{s}_{j}\cdot(\mathbf{r}_{ij}\times\mathbf{p}_{i})}{m_{i}m_{j}r_{ij}}\frac{\partial\tilde{G}^{so(v)}_{ij}}{\partial{r_{ij}}}\\
&&\equiv\tilde{H}^{so(v)ii}_{ij}+\tilde{H}^{so(v)jj}_{ij}+\tilde{H}^{so(v)ij}_{ij}.
\end{eqnarray}
The vectors $\mathbf{r}_{ij}$ and $\mathbf{p}_{i(j)}$ can be expressed in terms of the Jacobi coordinates ($\boldsymbol\rho$, $\boldsymbol\lambda$), by using the coordinates transformation,
\begin{eqnarray}
 \begin{aligned}
 &\mathbf{r}_{ij}=A_{rij}\boldsymbol\rho+B_{rij}\boldsymbol\lambda,\\
 &\mathbf{p}_{i}=A_{pi}\mathbf{p}_{\rho}+B_{pi}\mathbf{p}_{\lambda},
 \label{eq30}
 \end{aligned}
\end{eqnarray}
with $\boldsymbol\rho_{3}\equiv \boldsymbol\rho$ and $\boldsymbol\lambda_{3}\equiv \boldsymbol\lambda$. $A_{rij}$, $B_{rij}$, $A_{pi}$ and $B_{pi}$ can be obtained by Eq.~(\ref{e13}).  Taking the first part of $\tilde{H}^{so(v)}_{ij}$ as an example, we have
\begin{eqnarray}
 \begin{aligned}
 &\tilde{H}^{so(v)ii}_{ij}=\frac{\partial\tilde{G}^{so(v)}_{ii}}{r_{ij}\partial{r_{ij}}}[\frac{A_{rij}A_{pi}}{2m^{2}_{i}}\textbf{\emph{l}}_{\rho}\cdot\mathbf{s}_{i}
 +\frac{B_{rij}B_{pi}}{2m^{2}_{i}}\textbf{\emph{l}}_{\lambda}\cdot\mathbf{s}_{i}\\
  &+\frac{A_{rij}B_{pi}}{2m^{2}_{i}}(\boldsymbol\rho\times\mathbf{p}_{\lambda})\cdot\mathbf{s}_{i}
  +\frac{B_{rij}A_{pi}}{2m^{2}_{i}}(\boldsymbol\lambda\times\mathbf{p}_{\rho})\cdot\mathbf{s}_{i}].
 \label{eq31}
 \end{aligned}
\end{eqnarray}
The latter two terms proportional to $\boldsymbol\lambda\times\mathbf{p}_{\rho}$ or $\boldsymbol\rho\times\mathbf{p}_{\lambda}$ belong to the three-body spin-orbit potentials $\hat{H}^{so(v)-3B}_{ij}$. Their off-diagonal elements are considerably tougher. The difficulty lies in that the scalar variable $r_{ij}$ in the formula cannot be linearly expressed in terms of the Jacobi coordinates ($\boldsymbol\rho$, $\boldsymbol\lambda$), especially for $\hat{H}^{so(v)-3B}_{23}$ and $\hat{H}^{so(v)-3B}_{31}$.

Previously, Isgur et al. attempted to solve this problem by using the approximate conditions $\{\textbf{\emph{l}}_{\rho3},\textbf{\emph{l}}_{\lambda3}\}$ = $\{\textbf{\emph{l}}_{\rho1},\textbf{\emph{l}}_{\lambda1}\}$ = $\{\textbf{\emph{l}}_{\rho2},\textbf{\emph{l}}_{\lambda2}\}$. Here, 3, 1 and 2 denote the Jacobi coordinates JC-3, -1 and -2, respectively. Thus, $\hat{H}^{so(v)-3B}_{12}$, $\hat{H}^{so(v)-3B}_{23}$ and $\hat{H}^{so(v)-3B}_{31}$ were calculated in the Jacobi coordinates JC-3, -1 and -2, respectively. But, this approximation leads to two drawbacks: (1) The definition of orbital angular momenta $\{\textbf{\emph{l}}_{\rho},\textbf{\emph{l}}_{\lambda}\}$ is incorrect and confused. Actually, $\{\textbf{\emph{l}}_{\rho3},\textbf{\emph{l}}_{\lambda3}\}$, $\{\textbf{\emph{l}}_{\rho2},\textbf{\emph{l}}_{\lambda2}\}$ and  $\{\textbf{\emph{l}}_{\rho1},\textbf{\emph{l}}_{\lambda1}\}$ are different from each other. (2) The terms proportional to $\boldsymbol\lambda\times\mathbf{p}_{\rho}$ were lost in their calculations (see the Eq.(5.3) in~\cite{P39}, the Eqs.(B19) and (B20) in~\cite{P40} and the Eq.(43) in~\cite{P41}). So, the problem has not really been resolved. The rigorous calculation of the three-body spin-orbit potentials has always been a bottleneck for nearly 50 years.

In this work, we develop the GEM to the so-called two-step GEM to overcome this calculation problem.  Taking the last term in Eq.~(\ref{eq31}) as an example, its off-diagonal element is written as $\langle\alpha|\frac{\partial\tilde{G}^{so(v)}_{ii}}{r_{ij}\partial{r_{ij}}}\frac{B_{rij}A_{pi}}{2m^{2}_{i}}(\boldsymbol\lambda\times\mathbf{p}_{\rho})\cdot\mathbf{s}_{i}]|\beta\rangle$.
Expanding it using the Gaussian base functions as Eq.~(\ref{eq26}), we get as many as 10,000 matrix elements like $\langle(\tilde{n},\alpha)^{G}|\frac{\partial\tilde{G}^{so(v)}_{ii}}{r_{ij}\partial{r_{ij}}}\frac{B_{rij}A_{pi}}{2m^{2}_{i}}(\boldsymbol\lambda
\times\mathbf{p}_{\rho})\cdot\mathbf{s}_{i}]|(\tilde{n}',\beta)^{G}\rangle$, when $\tilde{n}_{max}$ =$100$.

We first deal with the right half part of the matrix element $[(\boldsymbol\lambda\times\mathbf{p}_{\rho})\cdot\mathbf{s}_{i}]|(\tilde{n}',\beta)^{G}\rangle$.
The vector operator product can be written in the component form,
\begin{eqnarray}
 \begin{aligned}
 (\boldsymbol\lambda\times\mathbf{p}_{\rho})\cdot\mathbf{s}_{i}=(\boldsymbol\lambda\times\mathbf{p}_{\rho})_{x}\cdot s_{ix}+(\boldsymbol\lambda\times\mathbf{p}_{\rho})_{y}\cdot s_{iy}\\+(\boldsymbol\lambda\times\mathbf{p}_{\rho})_{z}\cdot s_{iz},
 \end{aligned}
\end{eqnarray}
with
\begin{eqnarray}
 \begin{aligned}
 (\boldsymbol\lambda\times\mathbf{p}_{\rho})_{x}\cdot s_{ix}=\lambda_{y}p_{\rho z}s_{ix}-\lambda_{z}p_{\rho y}s_{ix},
 \end{aligned}
\end{eqnarray}
and the other two components can be written in a similar way. Then, the product breaks down into many small units  like $\lambda_{y}p_{\rho z}s_{ix}$, which facilitates the calculation of the right half part.

According to Eq.~(\ref{eq23}), when $\lambda_{y}p_{\rho z}s_{ix}$ acts on the orbital excited states, taking $i=1$ as an example, we have
\begin{eqnarray}
 \begin{aligned}
 &&(\lambda_{y}p_{\rho z}s_{1x})|(\tilde{n}',\beta)^{G}\rangle \\
 &&=(\lambda_{y}p_{\rho z}s_{1x})\sum_{\{m_{\xi}\}}\{CG_{\xi}\}\times|(n_{\rho}l_{\rho}m_{\rho})^{G} \rangle\otimes|(n_{\lambda}l_{\lambda}m_{\lambda})^{G} \rangle\\
&&\otimes|s_{1}m_{s_{1}}\rangle\otimes|s_{2}m_{s_{2}}\rangle\otimes|s_{3}m_{s_{3}}\rangle\\
&&=\sum_{\{m_{\xi}\}}\{CG_{\xi}\}\times(p_{\rho z}|(n_{\rho}l_{\rho}m_{\rho})^{G} \rangle)\times(\lambda_{y}|(n_{\lambda}l_{\lambda}m_{\lambda})^{G} \rangle)\\
&&\times(s_{1x}|s_{1}m_{s_{1}}\rangle)\times|s_{2}m_{s_{2}}\rangle\times|s_{3}m_{s_{3}}\rangle.
 \end{aligned}
\end{eqnarray}
The calculation of the terms $(p_{\rho z}|(n_{\rho}l_{\rho}m_{\rho})^{G} \rangle)$ and $(\lambda_{y}|(n_{\lambda}l_{\lambda}m_{\lambda})^{G} \rangle)$ can proceed by taking these operators as the spherical harmonic tensors, such as
\begin{eqnarray}
\begin{aligned}
&\lambda_{x}=\sqrt{\frac{2\pi}{3}}\lambda(Y_{1,-1}-Y_{1,1}), \\
&\lambda_{y}=i\sqrt{\frac{2\pi}{3}}\lambda(Y_{1,-1}+Y_{1,1}), \\
&\lambda_{x}=\sqrt{\frac{4\pi}{3}}\lambda Y_{1,0}.
\label{ee13}
\end{aligned}
\end{eqnarray}
By using the formula
\begin{eqnarray}
\begin{aligned}
&Y_{l_{1}m_{1}}Y_{l_{2}m_{2}}=\sum_{LM}\sqrt{\frac{(2l_{1}+1)(2l_{2}+1)}{4\pi(2L+1)}}C_{l_{1}0l_{2}0}^{L0}C_{l_{1}m_{1}l_{2}m_{2}}^{LM}Y_{LM},
\label{eq36}
\end{aligned}
\end{eqnarray}
the angular part of the function like $(\lambda_{y}|(n_{\lambda}l_{\lambda}m_{\lambda})^{G} \rangle)$ = $i\sqrt{\frac{2\pi}{3}}\lambda(|Y_{1,-1}\rangle+|Y_{1,1}\rangle)|\phi^{G}_{n_{\lambda}l_{\lambda}}(\lambda)Y_{l_{\lambda}m_{\lambda}}\rangle$ can be calculated
\begin{eqnarray}
\begin{aligned}
&y|(nlm)^{G} \rangle =(i\sqrt{\frac{2\pi}{3}})N_{nl}r^{l+1}e^{-\nu r^{2}}\sum_{l'm'} C_{l'm'}Y_{l'm'},
\label{eq37}
\end{aligned}
\end{eqnarray}
where, $C_{l'm'}$ is the abbreviation for all the coefficients in Eq.~(\ref{eq36}).
The resultant functions of Eq.~(\ref{eq37}) are generally not the standard Gaussian base functions. Once again, we expand them using the Gaussian base functions, for example,

\begin{eqnarray}
\begin{aligned}
&|r^{l+1}e^{-\nu r^{2}}Y_{l'm'}\rangle =\sum^{n'_{max}}_{n'=1}C'_{n'}|\phi^{G}_{n'l'm'}\rangle,
\label{eq38}
\end{aligned}
\end{eqnarray}
with
\begin{eqnarray}
\begin{aligned}
&C'_{n'} =\sum_{n''}(N^{-1})_{n'n''}\langle\phi^{G}_{n''l'm'}|r^{l+1}e^{-\nu r^{2}}Y_{l'm'}\rangle.
\label{ee113}
\end{aligned}
\end{eqnarray}

Similarly, $(p_{\rho z}|(n_{\rho}l_{\rho}m_{\rho})^{G} \rangle)$ can be calculated in the momentum space.   $[(\boldsymbol\lambda\times\mathbf{p}_{\rho})\cdot\mathbf{s}_{i}]|(\tilde{n}',\beta)^{G}\rangle$ can be then worked out. In this way, the calculation of matrix element $\langle(\tilde{n},\alpha)^{G}|\frac{\partial\tilde{G}^{so(v)}_{ii}}{r_{ij}\partial{r_{ij}}}\frac{B_{rij}A_{pi}}{2m^{2}_{i}}
(\boldsymbol\lambda\times\mathbf{p}_{\rho})\cdot\mathbf{s}_{i}]|(\tilde{n}',\beta)^{G}\rangle$ is reduced to computing a unit like
\begin{eqnarray}
 \begin{aligned}
 \notag
 \langle(\tilde{n},\alpha)^{G}|\frac{\partial\tilde{G}^{so(v)}_{ii}}{r_{ij}\partial{r_{ij}}}
 |(n'_{\rho}l'_{\rho}m'_{\rho})^{G} \rangle|(n'_{\lambda}l'_{\lambda}m'_{\lambda})^{G} \rangle
|s_{1}m'_{s_{1}}\rangle|s_{2}m'_{s_{2}}\rangle|s_{3}m'_{s_{3}}\rangle,
 \end{aligned}
\end{eqnarray}
if ignoring all of the coefficients involved, with the parity conservation condition $(-1)^{l'_{\rho}+l'_{\lambda}}=(-1)^{l_{\rho}+l_{\lambda}}$. At last,  $\frac{\partial\tilde{G}^{so(v)}_{ii}(r_{ij})}{r_{ij}\partial{r_{ij}}}$ (or $\frac{\partial\tilde{G}^{so(v)}_{ii}(\rho_{k})}{\rho_{k}\partial{\rho_{k}}}$) can be calculated by means of the infinitesimally-shifted Gaussian (ISG) basis functions~\cite{P47} and the coordinate transformation $(\boldsymbol\rho_{3}$, $\boldsymbol\lambda_{3})$ $\rightarrow$ $(\boldsymbol\rho_{k}$, $\boldsymbol\lambda_{k})$ ($k$ = 1, 2, 3).

The tensor terms can also be calculated in the same way, where the used spherical harmonic tensors formula is
\begin{eqnarray}
\begin{aligned}
&\hat{\textbf{s}}\cdot \hat{\textbf{r}}=\frac{\sqrt{2}}{2}(\hat{s}_{+}\hat{r}_{1,-1}+\hat{s}_{-}\hat{r}_{1,1})+\hat{s}_{z}\hat{r}_{1,0}.
\end{aligned}
\end{eqnarray}

In the calculations of the three-body spin-orbit terms and the tensor terms, there are over 4,000 computing units for each off-diagonal matrix element $\langle \alpha|\hat{O}|\beta\rangle/\sqrt{\langle\alpha|\alpha\rangle\langle\beta|\beta\rangle}$. By using an ordinary computer, all the rigorous calculations in this work can be completed within one week. The calculations of these off-diagonal matrix elements employ the GEM twice, that is the reason why we call the new method as the two-step GEM. Especially, the 2nd-step GEM in Eq.~(\ref{eq38}) is the key to working out those complicated matrix elements.

\subsection{Calculation results}

\begin{table*}[htbp]
\begin{ruledtabular}\caption{The Hamiltonian matrix $H$, the eigenvalues and the eigenvectors in the $\frac{1}{2}^{-}$, $\frac{3}{2}^{-}$ and $\frac{5}{2}^{-}$ subspaces of the charmed baryons. }
\begin{tabular}{c c c c c c c c c c c c c c c c c c c}
Baryon $(J^{P})$ &$(l_{\rho},l_{\lambda})nL(J^{P})_{j}$ & $H$ (MeV) & Eigenvalue (MeV) & Eigenvector   \\\hline
  \multirow{3}{*}{$\Sigma_{c}(\frac{1}{2}^{-})$} &$(0,1)1P(\frac{1}{2}^{-})_{0}$ & \multirow{3}{*}{$ \begin{pmatrix} 2773.06 & -11.33 & -3.31\\ -11.33 & 2778.02 & -22.26 \\ -3.31 & -22.26 & 2828.13\end{pmatrix} $}  &2836.60 & (0.0151428, -0.357707, 0.933711)\\
     & $(0,1)1P(\frac{1}{2}^{-})_{1}$ & &2783.20 &(0.757047, -0.605924, -0.244409)\\
    & $(1,0)1P(\frac{1}{2}^{-})_{1}$ & &2759.41 &(0.653185, 0.710564, 0.261626)\\\hline
    \multirow{3}{*}{$\Sigma_{c}(\frac{3}{2}^{-})$} &$(0,1)1P(\frac{3}{2}^{-})_{1}$ & \multirow{3}{*}{$ \begin{pmatrix} 2810.40 & 7.09 & -26.82\\ 7.09 & 2816.13 & -0.64 \\ -26.82 & -0.64 & 2877.37\end{pmatrix} $}  &2886.91 & (-0.333973, -0.041968, 0.941648)\\
     & $(0,1)1P(\frac{3}{2}^{-})_{2}$ & &2818.41 &(-0.31507, -0.936575, -0.153487)\\
    & $(1,0)1P(\frac{3}{2}^{-})_{1}$ & &2798.58 &(-0.888365, 0.347946, -0.299568)\\\hline
    $\Sigma_{c}(\frac{5}{2}^{-})$ & $(0,1)1P(\frac{5}{2}^{-})_{2}$ & (2862.97) & 2862.97 & (1) \\\hline
    \multirow{3}{*}{$\Xi'_{c}(\frac{1}{2}^{-})$} &$(0,1)1P(\frac{1}{2}^{-})_{0}$ & \multirow{3}{*}{$ \begin{pmatrix} 2906.46 & 10.77 & 3.81\\ 10.77 & 2911.67 & -13.56 \\ 3.81 & -13.56 & 2958.12\end{pmatrix} $}  &2961.80 & (0.0163397, -0.257791, 0.966062)\\
     & $(0,1)1P(\frac{1}{2}^{-})_{1}$ & &2918.64 &(0.682486, 0.708982, 0.177647)\\
    & $(1,0)1P(\frac{1}{2}^{-})_{1}$ & &2895.81 &(0.730716, -0.656421, -0.187523)\\\hline
    \multirow{3}{*}{$\Xi'_{c}(\frac{3}{2}^{-})$} &$(0,1)1P(\frac{3}{2}^{-})_{1}$ & \multirow{3}{*}{$ \begin{pmatrix} 2940.83 & 6.82 & 16.27\\ 6.82 & 2947.58 & 1.85 \\ 16.27 & 1.85 & 3004.10\end{pmatrix} $}  &3008.23 & (0.239675, 0.0565135, 0.969207)\\
     & $(0,1)1P(\frac{3}{2}^{-})_{2}$ & &2950.25 &(-0.395649, -0.905959, 0.150666)\\
    & $(1,0)1P(\frac{3}{2}^{-})_{1}$ & &2934.03 &(-0.886576, 0.419576, 0.194776)\\\hline
    $\Xi'_{c}(\frac{5}{2}^{-})$ & $(0,1)1P(\frac{5}{2}^{-})_{2}$ & (2990.24) & 2990.24 & (1) \\\hline
    \multirow{3}{*}{$\Omega_{c}(\frac{1}{2}^{-})$} &$(0,1)1P(\frac{1}{2}^{-})_{0}$ & \multirow{3}{*}{$ \begin{pmatrix} 3009.38 & -7.95 & 5.63\\ -7.95 & 3014.79 & 11.50 \\ 5.63 & 11.50 & 3058.84\end{pmatrix} $}  &3061.91 & (0.0700441, 0.225336, 0.97176)\\
     & $(0,1)1P(\frac{1}{2}^{-})_{1}$ & &3019.70 &(0.648008, -0.7509, 0.127414)\\
    & $(1,0)1P(\frac{1}{2}^{-})_{1}$ & &3001.40 &(-0.758406, -0.620784, 0.198616)\\\hline
    \multirow{3}{*}{$\Omega_{c}(\frac{3}{2}^{-})$} &$(0,1)1P(\frac{3}{2}^{-})_{1}$ & \multirow{3}{*}{$ \begin{pmatrix} 3044.57 & 4.50 & -12.94\\ 4.50 & 3051.82 & -4.38 \\ -12.94 & -4.38 & 3108.83\end{pmatrix} $}  &3111.78 & (-0.193897, -0.0859352, 0.977251)\\
     & $(0,1)1P(\frac{3}{2}^{-})_{2}$ & &3052.59 &(-0.299627, -0.943369, -0.142405)\\
    & $(1,0)1P(\frac{3}{2}^{-})_{1}$ & &3040.85 &(-0.934145, 0.320423, -0.157168)\\\hline
    $\Omega_{c}(\frac{5}{2}^{-})$ & $(0,1)1P(\frac{5}{2}^{-})_{2}$ & (3095.17) & 3095.17 & (1) \\
\end{tabular}
\end{ruledtabular}
\end{table*}

\begin{table*}[htbp]
\begin{ruledtabular}\caption{The Hamiltonian matrix $H$, the eigenvalues and the eigenvectors in the $\frac{1}{2}^{-}$, $\frac{3}{2}^{-}$ and $\frac{5}{2}^{-}$ subspaces of the bottom baryons. }
\begin{tabular}{c c c c c c c c c c c c c c c c c c c}
Baryon $(J^{P})$ &$(l_{\rho},l_{\lambda})nL(J^{P})_{j}$ & $H$ (MeV) & Eigenvalue (MeV) & Eigenvector   \\\hline
  \multirow{2}{*}{$\Sigma_{b}(\frac{1}{2}^{-})$} &$(0,1)1P(\frac{1}{2}^{-})_{0}$ & \multirow{2}{*}{$ \begin{pmatrix} 6087.11 & -5.08 \\ -5.08 & 6092.46 \end{pmatrix} $}  &6095.53 & (-0.516756, 0.856133)\\
     & $(0,1)1P(\frac{1}{2}^{-})_{1}$ & &6084.04 &(-0.856133, -0.516756)\\\hline
    \multirow{2}{*}{$\Sigma_{b}(\frac{3}{2}^{-})$} &$(0,1)1P(\frac{3}{2}^{-})_{1}$ & \multirow{2}{*}{$ \begin{pmatrix} 6105.46 & -3.33\\ -3.33 & 6113.18 \end{pmatrix} $}  &6114.42 & (-0.348442, 0.93733)\\
     & $(0,1)1P(\frac{3}{2}^{-})_{2}$ & &6104.22 &(-0.93733, -0.348442)\\\hline
    $\Sigma_{b}(\frac{5}{2}^{-})$ & $(0,1)1P(\frac{5}{2}^{-})_{2}$ & (6132.94) & 6132.94 & (1) \\\hline
  \multirow{2}{*}{$\Xi'_{b}(\frac{1}{2}^{-})$} &$(0,1)1P(\frac{1}{2}^{-})_{0}$ & \multirow{2}{*}{$ \begin{pmatrix} 6213.75 & 4.86 \\ 4.86 & 6218.47 \end{pmatrix} $}  &6221.51 & (0.530651, 0.84759)\\
     & $(0,1)1P(\frac{1}{2}^{-})_{1}$ & &6210.71 &(-0.84759, 0.530651)\\\hline
    \multirow{2}{*}{$\Xi'_{b}(\frac{3}{2}^{-})$} &$(0,1)1P(\frac{3}{2}^{-})_{1}$ & \multirow{2}{*}{$ \begin{pmatrix} 6230.38 & 3.31\\ 3.31 & 6237.41 \end{pmatrix} $}  &6238.72 & (0.36877, 0.929521)\\
     & $(0,1)1P(\frac{3}{2}^{-})_{2}$ & &6229.07 &(-0.929521, 0.36877)\\\hline
    $\Xi'_{b}(\frac{5}{2}^{-})$ & $(0,1)1P(\frac{5}{2}^{-})_{2}$ & (6255.74) & 6255.74 & (1) \\\hline
  \multirow{2}{*}{$\Omega_{b}(\frac{1}{2}^{-})$} &$(0,1)1P(\frac{1}{2}^{-})_{0}$ & \multirow{2}{*}{$ \begin{pmatrix} 6307.76 & -3.75 \\ -3.75 & 6313.35 \end{pmatrix} $}  &6315.23 & (-0.448552, 0.893757)\\
     & $(0,1)1P(\frac{1}{2}^{-})_{1}$ & &6305.88 &(-0.893757, -0.448552)\\\hline
    \multirow{2}{*}{$\Omega_{b}(\frac{3}{2}^{-})$} &$(0,1)1P(\frac{3}{2}^{-})_{1}$ & \multirow{2}{*}{$ \begin{pmatrix} 6325.67 & 2.18\\ 2.18 & 6334.46 \end{pmatrix} $}  &6334.97 & (0.2282, 0.973614)\\
     & $(0,1)1P(\frac{3}{2}^{-})_{2}$ & &6325.16 &(-0.973614, 0.2282)\\\hline
    $\Omega_{b}(\frac{5}{2}^{-})$ & $(0,1)1P(\frac{5}{2}^{-})_{2}$ & (6353.37) & 6353.37 & (1) \\
\end{tabular}
\end{ruledtabular}
\end{table*}

\begin{table*}[htbp]
\begin{ruledtabular}\caption{Hamiltonian expectation value $\langle H\rangle$ and contribution of each Hamiltonian term to the energy levels (in MeV) for the $1P$-wave states of the $\Sigma_{c}$ baryons with $\langle H^{mode}\rangle\equiv\langle H_{0}+H^{conf}\rangle$ and $\langle H_{ij}\rangle\equiv\langle H\rangle-\langle(H-H_{ij})\rangle$. The orbital excited states of the $\rho$-mode are marked in bold type. }
\begin{tabular}{c c c c c c c c c c c c c c c c c c c}
$(l_{\rho},l_{\lambda})nL(J^{P})_{j}$ & $\langle H^{mode}\rangle$ & $\{\langle H^{tens}_{12}\rangle$ & $\langle H^{tens}_{23}\rangle$ & $\langle H^{tens}_{31}\rangle\}$ & $\{\langle H^{cont}_{12}\rangle$  & $\langle H^{cont}_{23}\rangle$ & $\langle H^{cont}_{31}\rangle\}$ & $\{\langle H^{so(v)}_{12}\rangle$ & $\langle H^{so(v)}_{23}\rangle$ & $\langle H^{so(v)}_{31}\rangle\}$ & $\{\langle H^{so(s)}_{12}\rangle$ & $\langle H^{so(s)}_{23}\rangle$& $\langle H^{so(s)}_{31}\rangle\}$& $\langle H\rangle$  \\\hline
$(0,1)1P(\frac{1}{2}^{-})_{0}$ & 2781.78  &$\{$ 0 & 0 & 0 $\}$&$\{$ 42.06  & 0 & 0 $\}$&$\{$ 0 & -43.18 & -43.18 $\}$&$\{$ 0 & 17.15 & 17.15 $\}$& 2773.06  \\
$(0,1)1P(\frac{1}{2}^{-})_{1}$ & 2781.78  &$\{$ 0 & 0 & 0 $\}$&$\{$ 41.80  & -4.72 & -4.72 $\}$&$\{$ 0 & -29.27 & -29.27 $\}$&$\{$ 0 & 10.53 & 10.53 $\}$& 2778.02  \\
$(0,1)1P(\frac{3}{2}^{-})_{1}$ & 2781.78  &$\{$ 0 & 0.74 & 0.74 $\}$&$\{$ 41.13  & 2.14 & 2.14 $\}$&$\{$ 0 & -16.78 & -16.78 $\}$&$\{$ 0 & 7.79 & 7.79 $\}$& 2810.40  \\
$(0,1)1P(\frac{3}{2}^{-})_{2}$ & 2781.78  &$\{$ 0 & -0.44 & -0.44 $\}$&$\{$ 40.65  & -6.02 & -6.02 $\}$&$\{$ 0 & 9.38 & 9.38 $\}$&$\{$ 0 & -6.02 & -6.02 $\}$& 2816.13  \\
$\mathbf{(1,0)}1P(\frac{1}{2}^{-})_{1}$ & 2874.52 &$\{$ 0 & 0 & 0 $\}$&$\{$ -13.81  & 0 & 0 $\}$&$\{$ 0 & -16.64 & -16.64 $\}$&$\{$ 0 & 0 & 0 $\}$& \textbf{2828.13}  \\
$(0,1)1P(\frac{5}{2}^{-})_{2}$ & 2781.78  &$\{$ 0 & 1.38 & 1.38 $\}$&$\{$ 39.79  & 3.38 & 3.38 $\}$&$\{$ 0 & 25.01 & 25.01 $\}$&$\{$ 0 & -10.68 & -10.68 $\}$& 2862.97  \\
$\mathbf{(1,0)}1P(\frac{3}{2}^{-})_{1}$ & 2874.52 &$\{$ 0 & 0 & 0 $\}$&$\{$ -13.11  & 0 & 0 $\}$&$\{$ 0 & 8.07 & 8.07 $\}$&$\{$ 0 & 0 & 0 $\}$& \textbf{2877.37}  \\
\end{tabular}
\end{ruledtabular}
\end{table*}

\begin{table*}[htbp]
\begin{ruledtabular}\caption{Evolution of the energy levels (in MeV) for the $1P$-wave states of the $\Sigma_{c}$ baryons with successively adding the spin-dependent terms one by one. }
\begin{tabular}{c c c c c c c c c c c c c c c c c c c}
$(l_{\rho},l_{\lambda})nL(J^{P})_{j}$ & $\langle H^{mode}\rangle$  & $\langle\cdot+ H^{tens}\rangle$ & $\langle\cdot+ H^{cont}\rangle$  & $\langle\cdot+ H^{so(v)}\rangle$ & $\langle\cdot+ H^{so(s)}\rangle$ & $\langle H\rangle$  \\\hline
$(0,1)1P(\frac{1}{2}^{-})_{0}$ & 2781.78 & 2781.78 & 2822.52 & 2738.65 & 2773.06 & 2773.06  \\
$(0,1)1P(\frac{1}{2}^{-})_{1}$ & 2781.78 & 2781.78 &2814.34 &2756.94 &2778.02 & 2778.02  \\
$(0,1)1P(\frac{3}{2}^{-})_{1}$ & 2781.78 & 2783.30 &2827.94 &2794.79 &2810.40 & 2810.40  \\
$(0,1)1P(\frac{3}{2}^{-})_{2}$ & 2781.78 & 2780.86 &2809.24 &2828.12 &2816.13   & 2816.13  \\
$\mathbf{(1,0)}1P(\frac{1}{2}^{-})_{1}$ & 2874.52  &2874.52 &2861.19 &2828.13 &2828.13& \textbf{2828.13}  \\
$(0,1)1P(\frac{5}{2}^{-})_{2}$ & 2781.78 & 2784.82 &2833.28 &2884.30 &2862.97 & 2862.97  \\
$\mathbf{(1,0)}1P(\frac{3}{2}^{-})_{1}$ & 2874.52  & 2874.52 &2861.19 &2877.37 &2877.37 & \textbf{2877.37}  \\
\end{tabular}
\end{ruledtabular}
\end{table*}

\begin{table*}[htbp]
\begin{ruledtabular}\caption{Hamiltonian expectation value $\langle H\rangle$ and contribution of each Hamiltonian term to the energy levels (in MeV) for the $1P$-wave states of the $\Xi'_{c}$ baryons with $\langle H^{mode}\rangle\equiv\langle H_{0}+H^{conf}\rangle$ and $\langle H_{ij}\rangle\equiv\langle H\rangle-\langle(H-H_{ij})\rangle$. The orbital excited states of the $\rho$-mode are marked in bold type. }
\begin{tabular}{c c c c c c c c c c c c c c c c c c c}
$(l_{\rho},l_{\lambda})nL(J^{P})_{j}$ & $\langle H^{mode}\rangle$ & $\{\langle H^{tens}_{12}\rangle$ & $\langle H^{tens}_{23}\rangle$ & $\langle H^{tens}_{31}\rangle\}$ & $\{\langle H^{cont}_{12}\rangle$  & $\langle H^{cont}_{23}\rangle$ & $\langle H^{cont}_{31}\rangle\}$ & $\{\langle H^{so(v)}_{12}\rangle$ & $\langle H^{so(v)}_{23}\rangle$ & $\langle H^{so(v)}_{31}\rangle\}$ & $\{\langle H^{so(s)}_{12}\rangle$ & $\langle H^{so(s)}_{23}\rangle$& $\langle H^{so(s)}_{31}\rangle\}$& $\langle H\rangle$  \\\hline
$(0,1)1P(\frac{1}{2}^{-})_{0}$ & 2922.33  &$\{$ 0 & 0 & 0 $\}$&$\{$ 31.09  & 0 & 0 $\}$&$\{$ 0 & -34.13 & -41.02 $\}$&$\{$ 0 & 13.20 & 13.94 $\}$& 2906.46  \\
$(0,1)1P(\frac{1}{2}^{-})_{1}$ & 2922.33  &$\{$ 0 & 0 & 0 $\}$&$\{$ 30.94  & -3.11 & -5.06 $\}$&$\{$ 0 & -25.37 & -26.70 $\}$&$\{$ 0 & 8.58 & 8.90 $\}$& 2911.67  \\
$(0,1)1P(\frac{3}{2}^{-})_{1}$ & 2922.33  &$\{$ 0 & 0.89 & 0.57 $\}$&$\{$ 30.40  & 1.40 & 2.33 $\}$&$\{$ 0 & -12.12 & -16.55 $\}$&$\{$ 0 & 5.69 & 6.10 $\}$& 2940.83  \\
$(0,1)1P(\frac{3}{2}^{-})_{2}$ & 2922.33  &$\{$ 0 & -0.52 & -0.34 $\}$&$\{$ 30.14  & -3.96 & -6.63 $\}$&$\{$ 0 & 4.20 & 10.60 $\}$&$\{$ 0 & -3.82 & -4.27 $\}$& 2947.58  \\
$\mathbf{(1,0)}1P(\frac{1}{2}^{-})_{1}$ & 2999.89 &$\{$ 0 & 0 & 0 $\}$&$\{$ -11.32  & 0 & 0 $\}$&$\{$ 0 & -13.82 & -17.26 $\}$&$\{$ 0 & 0 & 0 $\}$& \textbf{2958.12}  \\
$(0,1)1P(\frac{5}{2}^{-})_{2}$ & 2922.33  &$\{$ 0 & 1.61 & 1.08 $\}$&$\{$ 29.45  & 2.18 & 3.82 $\}$&$\{$ 0 & 21.43 & 23.18 $\}$&$\{$ 0 & -8.52 & -8.88 $\}$& 2990.24  \\
$\mathbf{(1,0)}1P(\frac{3}{2}^{-})_{1}$ & 2999.89 &$\{$ 0 & 0 & 0 $\}$&$\{$ -10.68  & 0 & 0 $\}$&$\{$ 0 & 6.67 & 8.39 $\}$&$\{$ 0 & 0 & 0 $\}$& \textbf{3004.10}  \\
\end{tabular}
\end{ruledtabular}
\end{table*}

\begin{table*}[htbp]
\begin{ruledtabular}\caption{Evolution of the energy levels (in MeV) for the $1P$-wave states of the $\Xi'_{c}$ baryons with successively adding the spin-dependent terms one by one. }
\begin{tabular}{c c c c c c c c c c c c c c c c c c c}
$(l_{\rho},l_{\lambda})nL(J^{P})_{j}$ & $\langle H^{mode}\rangle$  & $\langle\cdot+ H^{tens}\rangle$ & $\langle\cdot+ H^{cont}\rangle$  & $\langle\cdot+ H^{so(v)}\rangle$ & $\langle\cdot+ H^{so(s)}\rangle$ & $\langle H\rangle$  \\\hline
$(0,1)1P(\frac{1}{2}^{-})_{0}$ &2922.33 &2922.33 &2952.48 &2879.29 &2906.46 &2906.46  \\
$(0,1)1P(\frac{1}{2}^{-})_{1}$ &2922.33 &2922.33 &2945.33 &2894.18 &2911.67  &2911.67 \\
$(0,1)1P(\frac{3}{2}^{-})_{1}$ &2922.33 &2923.81 &2957.39 &2929.03 &2940.83  &2940.83  \\
$(0,1)1P(\frac{3}{2}^{-})_{2}$ &2922.33 &2921.44 &2940.79 &2955.65 &2947.58  &2947.58  \\
$\mathbf{(1,0)}1P(\frac{1}{2}^{-})_{1}$ &2999.89 &2999.89 &2989.00 &2958.12 &2958.12 & \textbf{2958.12}  \\
$(0,1)1P(\frac{5}{2}^{-})_{2}$ & 2922.33 &2925.29 &2962.23 &3007.62 &2990.24 &2990.24 \\
$\mathbf{(1,0)}1P(\frac{3}{2}^{-})_{1}$ & 2999.89 &2999.89 &2989.00 &3004.10 &3004.10 & \textbf{3004.10}  \\
\end{tabular}
\end{ruledtabular}
\end{table*}

\begin{table*}[htbp]
\begin{ruledtabular}\caption{Hamiltonian expectation value $\langle H\rangle$ and contribution of each Hamiltonian term to the energy levels (in MeV) for the $1P$-wave states of the $\Omega_{c}$ baryons with $\langle H^{mode}\rangle\equiv\langle H_{0}+H^{conf}\rangle$ and $\langle H_{ij}\rangle\equiv\langle H\rangle-\langle(H-H_{ij})\rangle$. The orbital excited states of the $\rho$-mode are marked in bold type.  }
\begin{tabular}{c c c c c c c c c c c c c c c c c c c}
$(l_{\rho},l_{\lambda})nL(J^{P})_{j}$ & $\langle H^{mode}\rangle$ & $\{\langle H^{tens}_{12}\rangle$ & $\langle H^{tens}_{23}\rangle$ & $\langle H^{tens}_{31}\rangle\}$ & $\{\langle H^{cont}_{12}\rangle$  & $\langle H^{cont}_{23}\rangle$ & $\langle H^{cont}_{31}\rangle\}$ & $\{\langle H^{so(v)}_{12}\rangle$ & $\langle H^{so(v)}_{23}\rangle$ & $\langle H^{so(v)}_{31}\rangle\}$ & $\{\langle H^{so(s)}_{12}\rangle$ & $\langle H^{so(s)}_{23}\rangle$& $\langle H^{so(s)}_{31}\rangle\}$& $\langle H\rangle$  \\\hline
$(0,1)1P(\frac{1}{2}^{-})_{0}$ & 3032.90  &$\{$ 0 & 0 & 0 $\}$&$\{$ 24.43  & 0 & 0 $\}$&$\{$ 0 & -36.04 & -36.04 $\}$&$\{$ 0 & 11.43 & 11.43 $\}$& 3009.38  \\
$(0,1)1P(\frac{1}{2}^{-})_{1}$ & 3032.90  &$\{$ 0 & 0 & 0 $\}$&$\{$ 24.32  & -3.85 & -3.85 $\}$&$\{$ 0 & -25.84 & -25.84 $\}$&$\{$ 0 & 7.76 & 7.76 $\}$& 3014.79  \\
$(0,1)1P(\frac{3}{2}^{-})_{1}$ & 3032.90  &$\{$ 0 & 0.75 & 0.75 $\}$&$\{$ 23.87  & 1.73 & 1.73 $\}$&$\{$ 0 & -13.27 & -13.27 $\}$&$\{$ 0 & 4.75 & 4.75 $\}$& 3044.57  \\
$(0,1)1P(\frac{3}{2}^{-})_{2}$ & 3032.90  &$\{$ 0 & -0.44 & -0.44 $\}$&$\{$ 23.73  & -4.95 & -4.95 $\}$&$\{$ 0 & 5.85 & 5.85 $\}$&$\{$ 0 & -2.80 & -2.80 $\}$& 3051.82  \\
$\mathbf{(1,0)}1P(\frac{1}{2}^{-})_{1}$ & 3100.99 &$\{$ 0 & 0 & 0 $\}$&$\{$ -8.92  & 0 & 0 $\}$&$\{$ 0 & -16.97 & -16.97 $\}$&$\{$ 0 & 0 & 0 $\}$& \textbf{3058.84}  \\
$(0,1)1P(\frac{5}{2}^{-})_{2}$ & 3032.90  &$\{$ 0 & 1.38 & 1.38 $\}$&$\{$ 23.14  & 2.75 & 2.75 $\}$&$\{$ 0 & 21.81 & 21.81 $\}$&$\{$ 0 & -7.62 & -7.62 $\}$& 3095.17  \\
$\mathbf{(1,0)}1P(\frac{3}{2}^{-})_{1}$ & 3100.99 &$\{$ 0 & 0 & 0 $\}$&$\{$ -8.32  & 0 & 0 $\}$&$\{$ 0 & 8.14 & 8.14 $\}$&$\{$ 0 & 0 & 0 $\}$& \textbf{3108.83}  \\
\end{tabular}
\end{ruledtabular}
\end{table*}

\begin{table*}[htbp]
\begin{ruledtabular}\caption{Evolution of the energy levels (in MeV) for the $1P$-wave states of the $\Omega_{c}$ baryons with successively adding the spin-dependent terms one by one. }
\begin{tabular}{c c c c c c c c c c c c c c c c c c c}
$(l_{\rho},l_{\lambda})nL(J^{P})_{j}$ & $\langle H^{mode}\rangle$  & $\langle\cdot+ H^{tens}\rangle$ & $\langle\cdot+ H^{cont}\rangle$  & $\langle\cdot+ H^{so(v)}\rangle$ & $\langle\cdot+ H^{so(s)}\rangle$ & $\langle H\rangle$  \\\hline
$(0,1)1P(\frac{1}{2}^{-})_{0}$ &3032.90 &3032.90 &3056.60 &2986.51 &3009.38 &3009.38  \\
$(0,1)1P(\frac{1}{2}^{-})_{1}$ &3032.90 &3032.90 &3049.95 &2999.27 &3014.79  &3014.79 \\
$(0,1)1P(\frac{3}{2}^{-})_{1}$ &3032.90 &3034.42 &3061.30 &3035.07 &3044.57  &3044.57  \\
$(0,1)1P(\frac{3}{2}^{-})_{2}$ &3032.90 &3031.99 &3045.64 &3057.41 &3051.82  &3051.82\\
$\mathbf{(1,0)}1P(\frac{1}{2}^{-})_{1}$ &3100.99 &3100.99 &3092.48 &3058.84 &3058.84 & \textbf{3058.84}  \\
$(0,1)1P(\frac{5}{2}^{-})_{2}$ &3032.90 &3035.94 &3065.94 &3110.40 &3095.17 &3095.17 \\
$\mathbf{(1,0)}1P(\frac{3}{2}^{-})_{1}$ &3100.99 &3100.99 &3092.48 &3108.83 &3108.83 & \textbf{3108.83}  \\
\end{tabular}
\end{ruledtabular}
\end{table*}

\begin{table*}[htbp]
\begin{ruledtabular}\caption{Hamiltonian expectation value $\langle H\rangle$ and contribution of each Hamiltonian term to the energy levels (in MeV) for the $1P$-wave states of the $\Sigma_{b}$ baryons with $\langle H^{mode}\rangle\equiv\langle H_{0}+H^{conf}\rangle$ and $\langle H_{ij}\rangle\equiv\langle H\rangle-\langle(H-H_{ij})\rangle$. }
\begin{tabular}{c c c c c c c c c c c c c c c c c c c}
$(l_{\rho},l_{\lambda})nL(J^{P})_{j}$ & $\langle H^{mode}\rangle$ & $\{\langle H^{tens}_{12}\rangle$ & $\langle H^{tens}_{23}\rangle$ & $\langle H^{tens}_{31}\rangle\}$ & $\{\langle H^{cont}_{12}\rangle$  & $\langle H^{cont}_{23}\rangle$ & $\langle H^{cont}_{31}\rangle\}$ & $\{\langle H^{so(v)}_{12}\rangle$ & $\langle H^{so(v)}_{23}\rangle$ & $\langle H^{so(v)}_{31}\rangle\}$ & $\{\langle H^{so(s)}_{12}\rangle$ & $\langle H^{so(s)}_{23}\rangle$& $\langle H^{so(s)}_{31}\rangle\}$& $\langle H\rangle$  \\\hline
$(0,1)1P(\frac{1}{2}^{-})_{0}$ & 6072.01  &$\{$ 0 & 0 & 0 $\}$&$\{$ 42.56  & 0 & 0 $\}$&$\{$ 0 & -30.84 & -30.84 $\}$&$\{$ 0 & 17.05 & 17.05 $\}$& 6087.11  \\
$(0,1)1P(\frac{1}{2}^{-})_{1}$ & 6072.01  &$\{$ 0 & 0 & 0 $\}$&$\{$ 42.14  & -1.89 & -1.89 $\}$&$\{$ 0 & -17.78 & -17.78 $\}$&$\{$ 0 & 8.83 & 8.83 $\}$& 6092.46  \\
$(0,1)1P(\frac{3}{2}^{-})_{1}$ & 6072.01  &$\{$ 0 & 0.27 & 0.27 $\}$&$\{$ 41.82  & 0.91 & 0.91 $\}$&$\{$ 0 & -13.71 & -13.71 $\}$&$\{$ 0 & 8.51 & 8.51 $\}$& 6105.46  \\
$(0,1)1P(\frac{3}{2}^{-})_{2}$ & 6072.01  &$\{$ 0 & -0.16 & -0.16 $\}$&$\{$ 41.01  & -2.48 & -2.48 $\}$&$\{$ 0 & 10.97 & 10.97 $\}$&$\{$ 0 & -8.38 & -8.38 $\}$& 6113.18  \\
$(0,1)1P(\frac{5}{2}^{-})_{2}$ & 6072.01  &$\{$ 0 & 0.52 & 0.52 $\}$&$\{$ 40.53  & 1.54 & 1.54 $\}$&$\{$ 0 & 16.39 & 16.39 $\}$&$\{$ 0 & -9.03 & -9.03 $\}$& 6132.94  \\
\end{tabular}
\end{ruledtabular}
\end{table*}

\begin{table*}[htbp]
\begin{ruledtabular}\caption{Evolution of the energy levels (in MeV) for the $1P$-wave states of the $\Sigma_{b}$ baryons with successively adding the spin-dependent terms one by one. }
\begin{tabular}{c c c c c c c c c c c c c c c c c c c}
$(l_{\rho},l_{\lambda})nL(J^{P})_{j}$ & $\langle H^{mode}\rangle$  & $\langle\cdot+ H^{tens}\rangle$ & $\langle\cdot+ H^{cont}\rangle$  & $\langle\cdot+ H^{so(v)}\rangle$ & $\langle\cdot+ H^{so(s)}\rangle$ & $\langle H\rangle$  \\\hline
$(0,1)1P(\frac{1}{2}^{-})_{0}$ &6072.01 &6072.01 &6113.34 &6052.88 &6087.11 &6087.11  \\
$(0,1)1P(\frac{1}{2}^{-})_{1}$ &6072.01 &6072.01 &6109.89 &6074.77 &6092.46  &6092.46 \\
$(0,1)1P(\frac{3}{2}^{-})_{1}$ &6072.01 &6072.57 &6115.58 &6088.40 &6105.46  &6105.46  \\
$(0,1)1P(\frac{3}{2}^{-})_{2}$ &6072.01 &6071.67 &6107.82 &6129.92 &6113.18  &6113.18\\
$(0,1)1P(\frac{5}{2}^{-})_{2}$ &6072.01 &6073.13 &6117.80 &6150.98 &6132.94 &6132.94 \\
\end{tabular}
\end{ruledtabular}
\end{table*}

\begin{table*}[htbp]
\begin{ruledtabular}\caption{Hamiltonian expectation value $\langle H\rangle$ and contribution of each Hamiltonian term to the energy levels (in MeV) for the $1P$-wave states of the $\Xi'_{b}$ baryons with $\langle H^{mode}\rangle\equiv\langle H_{0}+H^{conf}\rangle$ and $\langle H_{ij}\rangle\equiv\langle H\rangle-\langle(H-H_{ij})\rangle$.  }
\begin{tabular}{c c c c c c c c c c c c c c c c c c c}
$(l_{\rho},l_{\lambda})nL(J^{P})_{j}$ & $\langle H^{mode}\rangle$ & $\{\langle H^{tens}_{12}\rangle$ & $\langle H^{tens}_{23}\rangle$ & $\langle H^{tens}_{31}\rangle\}$ & $\{\langle H^{cont}_{12}\rangle$  & $\langle H^{cont}_{23}\rangle$ & $\langle H^{cont}_{31}\rangle\}$ & $\{\langle H^{so(v)}_{12}\rangle$ & $\langle H^{so(v)}_{23}\rangle$ & $\langle H^{so(v)}_{31}\rangle\}$ & $\{\langle H^{so(s)}_{12}\rangle$ & $\langle H^{so(s)}_{23}\rangle$& $\langle H^{so(s)}_{31}\rangle\}$& $\langle H\rangle$  \\\hline
$(0,1)1P(\frac{1}{2}^{-})_{0}$ & 6206.92  &$\{$ 0 & 0 & 0 $\}$&$\{$ 31.43  & 0 & 0 $\}$&$\{$ 0 & -24.90 & -26.96 $\}$&$\{$ 0 & 13.18 & 13.97 $\}$& 6213.75  \\
$(0,1)1P(\frac{1}{2}^{-})_{1}$ & 6206.92  &$\{$ 0 & 0 & 0 $\}$&$\{$ 31.16  & -1.32 & -1.99 $\}$&$\{$ 0 & -15.20 & -15.26 $\}$&$\{$ 0 & 6.88 & 7.25 $\}$& 6218.47  \\
$(0,1)1P(\frac{3}{2}^{-})_{1}$ & 6206.92  &$\{$ 0 & 0.35 & 0.20 $\}$&$\{$ 30.91  & 0.63 & 0.96 $\}$&$\{$ 0 & -10.61 & -12.16 $\}$&$\{$ 0 & 6.50 & 6.91 $\}$& 6230.38  \\
$(0,1)1P(\frac{3}{2}^{-})_{2}$ & 6206.92  &$\{$ 0 & -0.21 & -0.12 $\}$&$\{$ 30.41  & -1.73 & -2.69 $\}$&$\{$ 0 & 7.61 & 10.09 $\}$&$\{$ 0 & -6.28 & -6.71 $\}$& 6237.41  \\
$(0,1)1P(\frac{5}{2}^{-})_{2}$ & 6206.92  &$\{$ 0 & 0.66 & 0.38 $\}$&$\{$ 30.04  & 1.07 & 1.69 $\}$&$\{$ 0 & 13.89 & 14.19 $\}$&$\{$ 0 & -6.94 & -7.32 $\}$& 6255.74  \\
\end{tabular}
\end{ruledtabular}
\end{table*}

\begin{table*}[htbp]
\begin{ruledtabular}\caption{Evolution of the energy levels (in MeV) for the $1P$-wave states of the $\Xi'_{b}$ baryons with successively adding the spin-dependent terms one by one. }
\begin{tabular}{c c c c c c c c c c c c c c c c c c c}
$(l_{\rho},l_{\lambda})nL(J^{P})_{j}$ & $\langle H^{mode}\rangle$  & $\langle\cdot+ H^{tens}\rangle$ & $\langle\cdot+ H^{cont}\rangle$  & $\langle\cdot+ H^{so(v)}\rangle$ & $\langle\cdot+ H^{so(s)}\rangle$ & $\langle H\rangle$  \\\hline
$(0,1)1P(\frac{1}{2}^{-})_{0}$ &6206.92 &6206.92 &6237.51 &6186.56 &6213.75 &6213.75  \\
$(0,1)1P(\frac{1}{2}^{-})_{1}$ &6206.92 &6206.92 &6234.46 &6204.33 &6218.47  &6218.47 \\
$(0,1)1P(\frac{3}{2}^{-})_{1}$ &6206.92 &6207.48 &6239.56 &6216.96 &6230.38  &6230.38  \\
$(0,1)1P(\frac{3}{2}^{-})_{2}$ &6206.92 &6206.58 &6232.59 &6250.39 &6237.41  &6237.41\\
$(0,1)1P(\frac{5}{2}^{-})_{2}$ &6206.92 &6208.05 &6241.60 &6269.98 &6255.74 &6255.74 \\
\end{tabular}
\end{ruledtabular}
\end{table*}

\begin{table*}[htbp]
\begin{ruledtabular}\caption{Hamiltonian expectation value $\langle H\rangle$ and contribution of each Hamiltonian term to the energy levels (in MeV) for the $1P$-wave states of the $\Omega_{b}$ baryons with $\langle H^{mode}\rangle\equiv\langle H_{0}+H^{conf}\rangle$ and $\langle H_{ij}\rangle\equiv\langle H\rangle-\langle(H-H_{ij})\rangle$. }
\begin{tabular}{c c c c c c c c c c c c c c c c c c c}
$(l_{\rho},l_{\lambda})nL(J^{P})_{j}$ & $\langle H^{mode}\rangle$ & $\{\langle H^{tens}_{12}\rangle$ & $\langle H^{tens}_{23}\rangle$ & $\langle H^{tens}_{31}\rangle\}$ & $\{\langle H^{cont}_{12}\rangle$  & $\langle H^{cont}_{23}\rangle$ & $\langle H^{cont}_{31}\rangle\}$ & $\{\langle H^{so(v)}_{12}\rangle$ & $\langle H^{so(v)}_{23}\rangle$ & $\langle H^{so(v)}_{31}\rangle\}$ & $\{\langle H^{so(s)}_{12}\rangle$ & $\langle H^{so(s)}_{23}\rangle$& $\langle H^{so(s)}_{31}\rangle\}$& $\langle H\rangle$  \\\hline
$(0,1)1P(\frac{1}{2}^{-})_{0}$ & 6309.57  &$\{$ 0 & 0 & 0 $\}$&$\{$ 24.86  & 0 & 0 $\}$&$\{$ 0 & -25.11 & -25.11 $\}$&$\{$ 0 & 11.63 & 11.63 $\}$& 6307.76  \\
$(0,1)1P(\frac{1}{2}^{-})_{1}$ & 6309.57  &$\{$ 0 & 0 & 0 $\}$&$\{$ 24.66  & -1.61 & -1.61 $\}$&$\{$ 0 & -15.01 & -15.01 $\}$&$\{$ 0 & 6.09 & 6.09 $\}$& 6313.35  \\
$(0,1)1P(\frac{3}{2}^{-})_{1}$ & 6309.57  &$\{$ 0 & 0.29 & 0.29 $\}$&$\{$ 24.44  & 0.77 & 0.77 $\}$&$\{$ 0 & -10.85 & -10.85 $\}$&$\{$ 0 & 5.70 & 5.70 $\}$& 6325.67  \\
$(0,1)1P(\frac{3}{2}^{-})_{2}$ & 6309.57  &$\{$ 0 & -0.17 & -0.17 $\}$&$\{$ 24.10  & -2.14 & -2.14 $\}$&$\{$ 0 & 8.13 & 8.13 $\}$&$\{$ 0 & -5.45 & -5.45 $\}$& 6334.46  \\
$(0,1)1P(\frac{5}{2}^{-})_{2}$ & 6309.57  &$\{$ 0 & 0.54 & 0.54 $\}$&$\{$ 23.78  & 1.33 & 1.33 $\}$&$\{$ 0 & 13.71 & 13.71 $\}$&$\{$ 0 & -6.10 & -6.10 $\}$& 6353.37  \\
\end{tabular}
\end{ruledtabular}
\end{table*}

\begin{table*}[htbp]
\begin{ruledtabular}\caption{Evolution of the energy levels (in MeV) for the $1P$-wave states of the $\Omega_{b}$ baryons with successively adding the spin-dependent terms one by one. }
\begin{tabular}{c c c c c c c c c c c c c c c c c c c}
$(l_{\rho},l_{\lambda})nL(J^{P})_{j}$ & $\langle H^{mode}\rangle$  & $\langle\cdot+ H^{tens}\rangle$ & $\langle\cdot+ H^{cont}\rangle$  & $\langle\cdot+ H^{so(v)}\rangle$ & $\langle\cdot+ H^{so(s)}\rangle$ & $\langle H\rangle$  \\\hline
$(0,1)1P(\frac{1}{2}^{-})_{0}$ &6309.57 &6309.57 &6333.78 &6284.49 &6307.76 &6307.76  \\
$(0,1)1P(\frac{1}{2}^{-})_{1}$ &6309.57 &6309.57 &6330.84 &6301.16 &6313.35 &6313.35 \\
$(0,1)1P(\frac{3}{2}^{-})_{1}$ &6309.57 &6310.15 &6335.80 &6314.27 &6325.67 &6325.67  \\
$(0,1)1P(\frac{3}{2}^{-})_{2}$ &6309.57 &6309.22 &6329.00 &6345.35 &6334.46  &6334.46\\
$(0,1)1P(\frac{5}{2}^{-})_{2}$ &6309.57 &6310.74 &6337.81 &6365.55 &6353.37 &6353.37 \\
\end{tabular}
\end{ruledtabular}
\end{table*}

\begin{table*}[htbp]
\begin{ruledtabular}\caption{The off-diagonal matrix elements (in MeV) of each Hamiltonian term for the negative-parity $\Sigma_{c}$ and $\Xi'_{c}$ baryon states. In the $\frac{1}{2}^{-}$ subspace, the three states are abbreviated as $|(0,1)1P(\frac{1}{2}^{-})_{0}\rangle\equiv|1\rangle$, $|(0,1)1P(\frac{1}{2}^{-})_{1}\rangle\equiv|2\rangle$ and $|(1,0)1P(\frac{1}{2}^{-})_{1}\rangle\equiv|3\rangle$, respectively. And in the $\frac{3}{2}^{-}$ subspace, the three states are labeled as $|(0,1)1P(\frac{3}{2}^{-})_{1}\rangle\equiv|1'\rangle$, $|(0,1)1P(\frac{3}{2}^{-})_{2}\rangle\equiv|2'\rangle$ and $|(1,0)1P(\frac{3}{2}^{-})_{1}\rangle\equiv|3'\rangle$, respectively. }
\begin{tabular}{c c c c c c c c c c c c c c c c c c c c c c c c c c c c c c}
\multirow{2}{*}{$\hat{O}$} & \multicolumn{3}{c}{$\Sigma_{c}(\frac{1}{2}^{-})$} & \multicolumn{3}{c}{$\Sigma_{c}(\frac{3}{2}^{-})$} & \multicolumn{3}{c}{$\Xi'_{c}(\frac{1}{2}^{-})$} & \multicolumn{3}{c}{$\Xi'_{c}(\frac{3}{2}^{-})$} \\ \cline{2-4} \cline{5-7} \cline{8-10} \cline{11-13}
   &$\langle1|\hat{O}|2\rangle$  & $\langle1|\hat{O}|3\rangle$  & $\langle2|\hat{O}|3\rangle$ & $\langle1'|\hat{O}|2'\rangle$  & $\langle1'|\hat{O}|3'\rangle$  & $\langle2'|\hat{O}|3'\rangle$ &$\langle1|\hat{O}|2\rangle$  & $\langle1|\hat{O}|3\rangle$  & $\langle2|\hat{O}|3\rangle$ & $\langle1'|\hat{O}|2'\rangle$  & $\langle1'|\hat{O}|3'\rangle$  & $\langle2'|\hat{O}|3'\rangle$  \\ \hline
$\hat{H}_{12}^{tens}$ & 0  & 0  & 0   & 0  & 0  & 0 & 0  & 0  & 0   & 0  & 0  & 0\\
$\hat{H}_{23}^{tens}$ &6.15 & -5.62 & -3.61 & -4.63  & 2.76 & -5.93     &-5.87 & 4.01 & -3.84 & -2.46  & -2.44 & 3.63\\
$\hat{H}_{31}^{tens}$ &6.15 & -5.62 & -3.61 & -4.63  & 2.76 & -5.93     &-6.34 & 6.48 & -3.10 & -6.76  & -2.84 & 7.72\\\hline
$\hat{H}_{12}^{cont}$ & 0  & 0  & 0   & 0  & 0  & 0 & 0  & 0  & 0   & 0  & 0  & 0\\
$\hat{H}_{23}^{cont}$ & -6.80& 3.96& 5.55 &4.65  &-2.57  & 5.61    & 4.48 & -2.96 & 4.14 &3.05  &1.89  & -4.15 \\
$\hat{H}_{31}^{cont}$ & -6.80& 3.96& 5.55 &4.65  &-2.57  & 5.61    & 7.28 & -3.72 & 5.22 &5.09  &2.44  & -5.36  \\\hline
$\hat{H}_{12}^{so(v)-2B}$ & 0  & 0  & 0   & 0  & 0  & 0  & 0  & 0  & 0   & 0  & 0  & 0\\
$\hat{H}_{23}^{so(v)-2B}$ & -11.33& 0& -7.06 &8.15& -6.87 & 0    & 12.28 & 0& -3.00 &8.80& 2.87 & 0\\
$\hat{H}_{31}^{so(v)-2B}$ & -11.33& 0& -7.06 &8.15& -6.87 & 0    & 9.06 & 0& -9.63 &6.59& 9.41 & 0\\\hline
$\hat{H}_{12}^{so(s)-2B}$ & 0  & 0  & 0   & 0  & 0  & 0  & 0  & 0  & 0   & 0  & 0  & 0\\
$\hat{H}_{23}^{so(s)-2B}$ & 2.73 & 0 & 1.82 & -2.08 & 1.84 & 0   & -2.79 & 0 & 0.68 & -2.13 & -0.68 & 0\\
$\hat{H}_{31}^{so(s)-2B}$ & 2.73 & 0 & 1.82 & -2.08 & 1.84 & 0   & -2.73 & 0 & 2.72 & -2.08 & -2.77 & 0\\\hline
$\hat{H}_{12}^{so(v)-3B}$ & 0 & 0 & -11.33 & 0 & -10.82 & 0 & 0 & 0 & -9.45 & 0 & 9.00 & 0  \\
$\hat{H}_{23}^{so(v)-3B}$ & 3.86& 0& 20.34 &-2.76& 19.48 & 0 & -2.77& 0& 12.58 &-1.96& -11.92 & 0 \\
$\hat{H}_{31}^{so(v)-3B}$ & 3.86 & 0 & 20.34 & -2.76 & 19.48 & 0 & -2.40 & 0 & 30.48 & -1.72 & -29.15 & 0\\\hline
$\hat{H}_{12}^{so(s)-3B}$ & 0& 0 & -19.19 & 0 &-19.00 & 0 & 0& 0 & -13.96 & 0 &13.76 & 0\\
$\hat{H}_{23}^{so(s)-3B}$ & -0.28& 0 & -12.91 & 0.20 & -13.15 & 0 & 0.17& 0 & -8.11 & 0.12 & 8.21 & 0\\
$\hat{H}_{31}^{so(s)-3B}$ & -0.28& 0 & -12.91 & 0.20 & -13.15 & 0 & 0.38& 0 & -18.30 & 0.28 & 18.48 & 0 \\\hline
Total & -11.33 & -3.31 & -22.26 & 7.09 & -26.82 & -0.64    & 10.77 & 3.81 & -13.56 & 6.82 & 16.27 & 1.85\\
\end{tabular}
\end{ruledtabular}
\end{table*}

\begin{table*}[htbp]
\begin{ruledtabular}\caption{The off-diagonal matrix elements (in MeV) of each Hamiltonian term for the negative-parity bottom baryon states. In the $\frac{1}{2}^{-}$ subspace, the two states are abbreviated as $|(0,1)1P(\frac{1}{2}^{-})_{0}\rangle\equiv|1\rangle$ and $|(0,1)1P(\frac{1}{2}^{-})_{1}\rangle\equiv|2\rangle$, respectively. In the $\frac{3}{2}^{-}$ subspace, the two states are abbreviated as $|(0,1)1P(\frac{3}{2}^{-})_{1}\rangle\equiv|1'\rangle$ and $|(0,1)1P(\frac{3}{2}^{-})_{2}\rangle\equiv|2'\rangle$, respectively. }
\begin{tabular}{c c c c c c c c c c c c c c c c c c c c c c c c c c c c c c}
\multirow{2}{*}{$\hat{O}$} &$\Sigma_{b}(\frac{1}{2}^{-})$  & $\Sigma_{b}(\frac{3}{2}^{-})$ & $\Xi'_{b}(\frac{1}{2}^{-})$ & $\Xi'_{b}(\frac{3}{2}^{-})$ & $\Omega_{b}(\frac{1}{2}^{-})$ & $\Omega_{b}(\frac{3}{2}^{-})$  \\ \cline{2-7}
   &$\langle1|\hat{O}|2\rangle$ &$\langle1'|\hat{O}|2'\rangle$&$\langle1|\hat{O}|2\rangle$&$\langle1'|\hat{O}|2'\rangle$&$\langle1|\hat{O}|2\rangle$&$\langle1'|\hat{O}|2'\rangle$ \\ \hline
$\hat{H}_{12}^{tens}$ & 0  & 0  & 0   & 0  & 0  & 0 \\
$\hat{H}_{23}^{tens}$ &2.27 & 1.80 & -2.34 & -1.08  & 2.52 & -2.09    \\
$\hat{H}_{31}^{tens}$ &2.27 & 1.80 & -2.26 & -2.49  & 2.52 & -2.09   \\\hline
$\hat{H}_{12}^{cont}$ & 0  & 0  & 0   & 0  & 0  & 0 \\
$\hat{H}_{23}^{cont}$ & -2.73& -1.94& 1.91 &1.35  &-2.32  & 1.66     \\
$\hat{H}_{31}^{cont}$ & -2.73& -1.94& 2.87 &2.08  &-2.32  & 1.66   \\\hline
$\hat{H}_{12}^{so(v)-2B}$ & 0  & 0  & 0   & 0  & 0  & 0 \\
$\hat{H}_{23}^{so(v)-2B}$ & -3.69& -2.72& 4.23 &3.11& -3.79 & 2.80   \\
$\hat{H}_{31}^{so(v)-2B}$ & -3.69& -2.72& 2.77 &2.07& -3.79 & 2.80   \\\hline
$\hat{H}_{12}^{so(s)-2B}$ & 0  & 0  & 0   & 0  & 0  & 0 \\
$\hat{H}_{23}^{so(s)-2B}$ & 0.35 & 0.27 & -0.38 & -0.29 & 0.38 & -0.29   \\
$\hat{H}_{31}^{so(s)-2B}$ & 0.35 & 0.27 & -0.34 & -0.26 & 0.38 & -0.29   \\\hline
$\hat{H}_{12}^{so(v)-3B}$ & 0 & 0 & 0 & 0 & 0 & 0  \\
$\hat{H}_{23}^{so(v)-3B}$ & 1.30& 0.96& -1.01 &-0.74& 1.37 & -1.01  \\
$\hat{H}_{31}^{so(v)-3B}$ & 1.30 & 0.96 & -0.66 & -0.49 & 1.37 & -1.01\\\hline
$\hat{H}_{12}^{so(s)-3B}$ & 0& 0 & 0 & 0 &0 & 0 \\
$\hat{H}_{23}^{so(s)-3B}$ & -0.04& -0.03 & 0.02 & 0.02 & -0.04  & 0.03\\
$\hat{H}_{31}^{so(s)-3B}$ & -0.04& -0.03 & 0.05 & 0.04 & -0.04 & 0.03  \\\hline
Total & -5.08 & -3.33 & 4.86 & 3.31 & -3.75 & 2.18    \\
\end{tabular}
\end{ruledtabular}
\end{table*}

\begin{table*}[htbp]
\begin{ruledtabular}\caption{The deviations of the calculated masses of the 22 baryons from the measured ones~\cite{F201,F210}. Most of the deviations are less than 8 MeV. The arithmetic average deviation $(\sum_{i=1}^{n}|M_{cal.}-M_{exp.}|_{i})/n$ is about \textbf{4.58} MeV. $M_{exp.}$ denotes the central value of the measured mass. All the masses are measured in MeV. }
\begin{tabular}{c c c c c | c c c c c }
\label{tb19}
Baryon ($J^{P}$) & $M_{exp.}$  & Predicted state & $M_{cal.}$ & $M_{cal.}$-$M_{exp.}$  & Baryon ($J^{P}$) & $M_{exp.}$  & Predicted state & $M_{cal.}$ & $M_{cal.}$-$M_{exp.}$ \\ \hline
$\Sigma_{c}(2800)^{++}(?^{?})$ & 2801  & $\Sigma_{c}(2799)\frac{3}{2}^{-}$ & 2798.58 & -2.42 & $\Omega_{c}(3000)^{0}(?^{?})$ & 3000.46 & $\Omega_{c}(3001)\frac{1}{2}^{-}$& 3001.40 & 0.94 \\
$\Sigma_{c}(2800)^{+}(?^{?})$ & 2792   & $\Sigma_{c}(2799)\frac{3}{2}^{-}$ & 2798.58 & 6.58& $\Omega_{c}(3050)^{0}(?^{?})$ & 3050.17  & $\Omega_{c}(3053)\frac{3}{2}^{-}$ & 3052.59 & 2.42  \\
$\Sigma_{c}(2800)^{0}(?^{?})$ & 2806  & $\Sigma_{c}(2799)\frac{3}{2}^{-}$ & 2798.58 & -7.42 & $\Omega_{c}(3065)^{0}(?^{?})$ & 3065.58  & $\Omega_{c}(3062)\frac{1}{2}^{-}$ & 3061.91 & -3.67  \\
$\Sigma_{c}(2846)^{0}(?^{?})$ & 2846  & $\Sigma_{c}(2837)\frac{1}{2}^{-}$ & 2836.60 & -9.40 & $\Omega_{c}(3090)^{0}(?^{?})$ & 3090.15 & $\Omega_{c}(3095)\frac{5}{2}^{-}$ & 3095.17 & 5.02  \\
$\Xi_{c}(2882)^{0}(?^{?})$ & 2882  & $\Xi'_{c}(2896)\frac{1}{2}^{-}$ & 2895.81 & 13.81 & $\Omega_{c}(3120)^{0}(?^{?})$ & 3118.98  & $\Omega_{c}(3112)\frac{3}{2}^{-}$ & 3111.78 & -7.91  \\
$\Xi_{c}(2923)^{+}(?^{?})$ & 2922.8  & $\Xi'_{c}(2919)\frac{1}{2}^{-}$ & 2918.64 & -4.16 & $\Omega_{b}(6316)^{-}(?^{?})$  & 6315.6 & $\Omega_{b}(6315)\frac{1}{2}^{-}$ & 6315.23 & -0.37  \\
$\Xi_{c}(2923)^{0}(?^{?})$ & 2923.2  & $\Xi'_{c}(2919)\frac{1}{2}^{-}$ & 2918.64 & -4.56 & $\Omega_{b}(6330)^{-}(?^{?})$ & 6330.3 & $\Omega_{b}(6325)\frac{3}{2}^{-}$ & 6325.16 & -5.14  \\
$\Xi_{c}(2930)^{+}(?^{?})$ & 2942   & $\Xi'_{c}(2934)\frac{3}{2}^{-}$ & 2934.03 &-7.97 & $\Omega_{b}(6340)^{-}(?^{?})$  & 6339.7 & $\Omega_{b}(6335)\frac{3}{2}^{-}$ & 6334.97 & -4.73  \\
$\Xi_{c}(2930)^{0}(?^{?})$ & 2938.55  & $\Xi'_{c}(2934)\frac{3}{2}^{-}$ & 2934.03 & -4.52& $\Omega_{b}(6350)^{-}(?^{?})$ & 6349.8  & $\Omega_{b}(6353)\frac{5}{2}^{-}$ & 6353.37 & 3.57 \\
$\Sigma_{b}(6097)^{+}(?^{?})$ & 6095.8& $\Sigma_{b}(6096)\frac{1}{2}^{-}$  & 6095.53  & -0.27 &$\Xi_{b}(6227)^{0}(?^{?})$ & 6226.8 & $\Xi'_{b}(6229)\frac{3}{2}^{-}$ & 6229.07 & 2.27 \\
$\Sigma_{b}(6097)^{-}(?^{?})$ & 6098.0  & $\Sigma_{b}(6096)\frac{1}{2}^{-}$ & 6095.53 & -2.47 & $\Xi_{b}(6227)^{-}(?^{?})$  & 6227.9  & $\Xi'_{b}(6229)\frac{3}{2}^{-}$ & 6229.07 & 1.17 \\
\end{tabular}
\end{ruledtabular}
\end{table*}


\begin{thebibliography}{}

\bibitem{F101}
F.~Gross, E.~Klempt, S.~J.~Brodsky, A.~J.~Buras, V.~D.~Burkert et al., 50 Years of Quantum Chromodynamics,
\href{https://doi.org/10.1140/epjc/s10052-023-11949-2}{Eur.Phys.J.C \textbf{83}, 1125 (2023)}, arXiv:2212.11107[hep-ph].

\bibitem{F201}S. Navas, et al., (Particle Data Group), Review of Particle Physics,
\href{https://doi.org/10.1103/PhysRevD.110.030001}{Phys. Rev. D \textbf{110} 3, 030001,(2024) and 2025 update}.

\bibitem{F203}R. Aaij, et al., (LHCb collaboration), Observation of excited $\Omega_{c}^{0}$ baryons in $\Omega_{b}^{-}\rightarrow \Xi_{c}^{+}K^{-}\pi^{-}$ decays,
\href{https://doi.org/10.1103/PhysRevD.104.L091102}{Phys. Rev. D \textbf{104} 9 , L091102 (2021)}, arXiv:2107.03419[hep-ex].

\bibitem{F204}R. Aaij, et al., (LHCb collaboration), Observation of New $\Omega_{c}^{0}$ baryons Decay to $\Lambda_{c}^{+}K^{-}$,
\href{https://doi.org/10.1103/PhysRevLett.124.222001}{Phys. Rev. Lett. \textbf{124}, 222001 (2020)}, arXiv:2003.13649[hep-ex].

\bibitem{F205}Y. B. Li, et al., (Belle  Collaboration), Evidence of a structure in $\bar{K}^{0}\Lambda_{c}^{+}$ consistant with a charged $\Xi_{c}(2930)^{+}$, and updated measurement of $\bar{B}^{0}\rightarrow \bar{K}^{0}\Lambda_{c}^{+}\bar{\Lambda}_{c}^{-}$ at Belle,
\href{https://doi.org/10.1140/epjc/s10052-018-6425-5}{Eur. Phys. J. C \textbf{78}, 928 (2018)}, arXiv:1806.09182[hep-ex].

\bibitem{F206}B. Aubert, et al., (BABAR Collaboration), A study of Excited Charm-Strange Baryons with Evidence for new Baryons $\Xi_{c}(3055)^{+}$ and $\Xi_{c}(3123)^{+}$,
\href{https://doi.org/10.1103/PhysRevD.77.012002}{Phys. Rev. D \textbf{77}, 012002 (2008)}, arXiv:0710.5763[hep-ex].

\bibitem{F207}R. Aaij, et al.,(LHCb collaboration), Observation of New $\Omega_{c}^{0}$ States Decaying to the $\Xi_{c}^{+}K^{-}$ Final State,
\href{https://doi.org/10.1103/PhysRevLett.131.131902}{Phys. Rev. Lett. \textbf{131} 13, 131902 (2023)}, arXiv:2302.04733[hep-ex].

\bibitem{F208}R. Aaij, et al., (LHCb collaboration), Observation of New Baryons in the $\Xi_{b}^{-}\pi^{+}\pi^{-}$ and $\Xi_{b}^{0}\pi^{+}\pi^{-}$ Systems,
\href{https://doi.org/10.1103/PhysRevLett.131.171901}{Phys. Rev. Lett. \textbf{131} 17, 171901 (2023)}, arXiv:2307.13399[hep-ex].

\bibitem{F209}Y. B. Li, et al., (Belle  Collaboration), Observation of $\Xi_{c}(2930)^{0}$ and updated measurement of $B^{-}\rightarrow K^{-}\Lambda_{c}^{+}\bar{\Lambda}_{c}^{-}$ at Belle,
\href{https://doi.org/10.1140/epjc/s10052-018-5720-5}{Eur. Phys. J. C \textbf{78} 3, 252 (2018)}, arXiv:1712.03612[hep-ex].

\bibitem{F210}B. Aubert, et al., (BABAR  Collaboration), Measurements of $\mathcal{B}(\bar{B}^{0}\rightarrow\Lambda_{c}^{+}\bar{p})$ and $\mathcal{B}(\bar{B}^{-}\rightarrow\Lambda_{c}^{+}\bar{p}\pi^{-})$ and Studies of $\Lambda_{c}^{+}\pi^{-}$ Resonances,
\href{https://doi.org/10.1103/PhysRevD.78.112003}{Phys. Rev. D \textbf{78}, 112003 (2008)}, arXiv:0807.4974[hep-ex].



\bibitem{P02}M. Karliner and J. L. Rosner, Very narrow excited $\Omega_{c}$ baryons,
\href{https://doi.org/10.1103/PhysRevD.95.114012}{Phys. Rev. D \textbf{95} 11, 114012 (2017)}, arXiv:1703.07774[hep-ph].

\bibitem{P03}E. Ortiz-Pacheco, R. Bijker, A. Giachino, and E. Santopinto, Heavy $\Omega_{c}$ and $\Omega_{b}$  baryons in the quark model,
\href{https://doi.org/10.1088/1742-6596/1610/1/012011}{J. Phys. Conf. Ser. \textbf{1610} 1, 012011 (2020)}, arXiv:2004.09409[nucl-th].

\bibitem{P05}Z. Shah, K. Thakkar, A. K. Rai, and P. C. Vinodkumar, Excited State Mass spectra of Singly Charmed Baryons,
\href{https://doi.org/10.1140/epja/i2016-16313-9}{Eur. Phys. J. A \textbf{52}, 313 (2016)}, arXiv:1602.06384[hep-ph].

\bibitem{P06}A. Kakadiya, Z. Shah, and A. K. Rai, Mass spectra and decay properties of singly heavy bottom-strange baryons,
\href{https://doi.org/10.1142/S0217751X22500531}{Int. J. Mod. Phys. A \textbf{37} 11n12, 2250053 (2022)}, arXiv: 2202.12048 [hep-ph].

\bibitem{P07}K. L. Wang,  L. Y. Xiao, X. H. Zhong, and Q. Zhao, Understanding the newly observed $\Omega_{c}$ states through their decays,
\href{https://doi.org/10.1103/PhysRevD.95.116010}{Phys. Rev. D \textbf{95} 11, 116010 (2017)}, arXiv:1703.09130[hep-ph].

\bibitem{P08}K. L. Wang, Q. F. L\"{u}, and X. H. Zhong, Interpretation of the newly observed $\Sigma_{b}(6097)^{\pm}$ and $\Xi_{b}(6227)^{-}$ states as the $P$-wave bottom baryons,
\href{https://doi.org/10.1103/PhysRevD.99.014011}{Phys. Rev. D \textbf{99} 1, 014011 (2019)}, arXiv:1810.02205[hep-ph].

\bibitem{P09}Z. G. Wang, Analysis of $\Omega_{c}(3000)$, $\Omega_{c}(3050)$, $\Omega_{c}(3066)$, $\Omega_{c}(3090)$ and $\Omega_{c}(3119)$ with QCD sum rules,
\href{https://doi.org/10.1140/epjc/s10052-017-4895-5}{Eur. Phys. J. C \textbf{77} 5, 325 (2017)}, arXiv:1704.01854[hep-ph].

\bibitem{P10}S. S. Agaev, K. Azizi, and H. Sundu, Interpretation of the new $\Omega_{c}^{0}$ states via their mass and width,
\href{https://doi.org/10.1140/epjc/s10052-017-4953-z}{Eur. Phys. J. C \textbf{77} 6, 395 (2017)}, arXiv:1704.04928[hep-ph].

\bibitem{P11}Z. G. Wang, Analysis of $\Omega_{b}(6316)$, $\Omega_{b}(6330)$, $\Omega_{b}(6340)$ and $\Omega_{b}(6350)$ with QCD sum rules,
\href{https://doi.org/10.1142/S0217751X20500438}{Int. J. Mod. Phys. A \textbf{35} 07, 2050043 (2020)}, arXiv:2001.02961[hep-ph].

\bibitem{P12}H. M. Yang and H. X Chen, $P$-wave charmed baryons of the $SU(3)$ flavor $6_{F}$,
\href{https://doi.org/10.1103/PhysRevD.104.034037}{Phys. Rev. D \textbf{104} 3, 034037 (2021)}, arXiv:2106.15488[hep-ph].

\bibitem{P01}W. Roberts and M. Pervin, Heavy baryons in a quark model,
\href{https://doi.org/10.1142/S0217751X08041219}{Int. J. Mod. Phys. A \textbf{23}, 2817 (2008)}, arXiv:0711.2492[nucl-th].

\bibitem{P13}D. Ebert, R. N. Faustov, and V. O. Galkin, Spectroscopy and Regge trajectories of heavy baryons in the relativistic quark-diquark picture,
\href{https://doi.org/10.1103/PhysRevD.84.014025}{Phys. Rev. D \textbf{84}, 014025 (2011)}, arXiv:1105.0583[hep-ph].

\bibitem{P14}B. Chen and X. Liu, Assigning the newly reported $\Sigma_{b}(6097)$ as a $P$-wave excited state and predicting its partners,
\href{https://doi.org/10.1103/PhysRevD.98.074032}{Phys. Rev. D \textbf{98}, 074032 (2018)}, arXiv:1810.00389[hep-ph].

\bibitem{P141}B. Chen, K. W. Wei, X. Liu, and T. Matsuki, Low-lying charmed and charmed-strange baryon states,
\href{https://doi.org/10.1140/epjc/s10052-017-4708-x}{Eur. Phys. J. C \textbf{77}, 154 (2017)}, arXiv:1609.07967[hep-ph].

\bibitem{P15}M. Padmanath and N. Mathur, Quantum Numbers of Recently Discovered $\Omega_{c}^{0}$  Baryons from Lattice QCD,
\href{https://doi.org/10.1103/PhysRevLett.119.042001}{Phys. Rev. Lett. \textbf{119} 4, 042001 (2017)}, arXiv:1704.00259[hep-ph].

\bibitem{P16}Z. Zhao, D. D. Ye, and A. Zhang, Hadronic decay properties of newly observed $\Omega_{c}$ baryons,
\href{https://doi.org/10.1103/PhysRevD.95.114024}{Phys. Rev. D \textbf{95} 11, 114024 (2017)}, arXiv:1704.02688[hep-ph].

\bibitem{P17}R. Bijker, H. Garc\'{\i}a-Tecocoatzi and A. Giachino, et al., Masses and decay widths of $\Xi_{_{c/b}}$ and $\Xi_{c/b}^{'}$ baryons,
\href{https://doi.org/10.1103/PhysRevD.105.074029}{Phys. Rev. D \textbf{105}, 074029 (2022)}, arXiv:2010.12437[hep-ph].

\bibitem{P181}Q. Xin, Z. G. Wang and F. L\"{u}, The $\Lambda$-type $P$-wave bottom baryon states via the QCD sum rules,
\href{https://doi.org/10.1088/1674-1137/ace81f}{Chin. Phys. C \textbf{47}, 093106 (2023)}, arXiv:2306.05626[hep-ph].

\bibitem{P18}H. Garc\'{\i}a-Tecocoatzi, A. Giachino, and A. Ramirez-Morales, et al., Decay widths and mass spectra of single bottom baryons,
\href{https://doi.org/10.48550/arXiv.2307.00505}{arXiv:2307.00505[hep-ph]}.

\bibitem{S01}E, Willis, J. Lamb, and R. C. Retherford, Fine structure of the hydrogen atom by a microwave method, Phys. Rev. 1947; 72: 241-3.
\bibitem{S02}M. G. Mayer, On Closed Shells in Nuclei. II, Phys. Rev. 75, 1969 (1949).

\bibitem{P20}H. Bahtiyar, K. U. Can, G. Erkol, P. Gubler, and M. Oka, Charmed baryon spectrum from lattice QCD near the physical point,
\href{https://doi.org/10.1103/PhysRevD.102.054513}{Phys. Rev. D \textbf{102} 5, 054513 (2020)}, arXiv:2004.08999[hep-lat].

\bibitem{P21}Z. G. Wang, Review of the QCD sum rules for exotic states,
\href{https://doi.org/10.15302/frontphys.2026.016300}{Front. Phys. (Beijing) \textbf{21} 1, 016300 (2026)}, arXiv:2502.11351[hep-ph].

\bibitem{P22}X. Luo, S. W. Zhang, H. X. Chen, and A. Hosaka, et al., A short review on QCD sum rule studies of $P$-wave single heavy baryons,
\href{https://doi.org/10.1007/s43673-026-00188-8}{AAPPS Bull. \textbf{36} 1, 9 (2026)}, arXiv:2510.13013[hep-ph].

\bibitem{P221}T. M. Aliev, K. Azizi, Y. Sarac, and H. Sundu, Determination of the quantum numbers of $\Sigma_{b}(6097)^{\pm}$ via their strong decays,
\href{https://doi.org/10.1103/PhysRevD.99.094003}{Phys. Rev. D \textbf{99} 9, 094003 (2019)}, arXiv:1811.05686[hep-ph].

\bibitem{P23}H. Y. Cheng, and C. K. Chua, Strong Decays of Charmed Baryons in Heavy Hadron Chiral Perturbation Theory: An Update,
\href{https://doi.org/10.1103/PhysRevD.92.074014}{Phys. Rev. D \textbf{92} 7, 074014 (2015)}, arXiv:1508.05653[hep-ph].



\bibitem{P24}J. X. Lu, Y. Zhou, and H. X. Chen, et al., Dynamically generated $J^{P}$ = $1/2^{-}$ ($3/2^{-}$)singly charmed and bottom heavy baryons,
\href{https://doi.org/10.1103/PhysRevD.92.014036}{Phys. Rev. D \textbf{92} 1, 014036 (2015)}, arXiv:1409.3133[hep-ph].

\bibitem{P251}H. X. Chen, W. Chen, and X. Liu, et al., A review of the open charm and open bottom systems,
\href{https://doi.org/10.1088/1361-6633/aa6420}{Rept. Prog. Phys. \textbf{80} 7, 076201 (2017)}, arXiv:1609.08928[hep-ph].

\bibitem{P25}D. J. Jia, W. N. Liu, and A. Hosaka,  Regge behaviors in orbitally excited spectroscopy of charmed and bottom baryons,
\href{https://doi.org/10.1103/PhysRevD.92.014036}{Phys. Rev. D \textbf{101} 3, 034016 (2020)}, arXiv:1907.04958[hep-ph].

\bibitem{P261}L. A. Copley, N. Isgur and G. Karl, Charmed Baryons in a Quark Model with Hyperfine Interactions,
\href{https://doi.org/10.1103/PhysRevD.20.768, 10.1103/PhysRevD.23.817.3}{Phys. Rev. D \textbf{20}, 768 (1979)},{~Erratum: [Phys. Rev. D \textbf{23}, 817 (1981)]}.

\bibitem{P26}S. Tawfiq, P. J. O'Donnell, and J. G. Korner, Charmed baryon strong coupling constants in a light front quark  model,
\href{https://doi.org/10.1103/PhysRevD.58.054010}{Phys. Rev. D \textbf{58}, 054010 (1998)}, arXiv:hep-ph/9803246.

\bibitem{P27}M. A. Ivanov, J. G. Korner, V. E. Lyubovitskij, and A. G. Rusetsky, One pion charm baryon transitions  in a relativistic three quark model,
\href{https://doi.org/10.1016/S0370-2693(98)01245-3}{Phys. Lett. B \textbf{442}, 435-442 (1998)}, arXiv:hep-ph/9807519.

\bibitem{P28}C. W. Hwang, Combined Chiral Dynamics and MIT Bag Model Study of Strong $\Sigma*_{Q}\rightarrow \Lambda_{Q}\pi$ Decay,
\href{https://doi.org/10.1140/epjc/s10052-007-0267-x}{Eur. Phys. J. C \textbf{50}, 793-799 (2007)}, arXiv:hep-ph/0611221.

\bibitem{P29}D. Ebert, R. N. Faustov, and V. O. Galkin, Masses of excited heavy baryons in the relativistic quark model,
\href{https://doi.org/10.1016/j.physletb.2007.11.037}{Phys. Lett. B \textbf{659}, 612-620 (2008)}, arXiv:0705.2957.


\bibitem{P31}X. H. Zhong, and Q. Zhao, Charmed baryon strong decays in a chiral quark model,
\href{https://doi.org/10.1103/PhysRevD.77.074008}{Phys. Rev. D \textbf{77}, 074008 (2008)}, arXiv:0711.4645.

\bibitem{P32}T. Yoshida, E. Hiyama, and A. Hosaka, et al., Spectrum of heavy baryons in the quark model,
\href{https://doi.org/10.1103/PhysRevD.92.114029}{Phys. Rev. D \textbf{92}, 114029 (2015)}, arXiv:1510.01067[hep-ph].

\bibitem{P33}Q. F. L\"{u}, K. L. Wang, L. Y. Xiao, and X. H. Zhong, Mass spectra and radiative transitions of doubly heavy baryons in a relativized quark model,
\href{https://doi.org/10.1103/PhysRevD.96.114006}{Phys. Rev. D \textbf{96} 11, 114006 (2017)}, arXiv:1708.04468[hep-ph].


\bibitem{P34}B. Chen, S. Q. Luo, and X. Liu, Universal behavior of mass gaps existing in the single heavy baryon family,
\href{https://doi.org/10.1103/PhysRevD.96.114006}{Eur. Phys. J. C \textbf{81} 5, 474 (2021)}, arXiv:2101.10806[hep-ph].

\bibitem{P35}Y. X. Peng, S. Q. Luo, and X. Liu, Refining radiative decay studies in singly heavy baryons,
\href{https://doi.org/10.1103/PhysRevD.110.074034}{Phys. Rev. D \textbf{110} 7, 074034 (2024)}, arXiv:2405.12812[hep-ph].

\bibitem{P36}X. Z. Weng, W. Z. Deng and S. L. Zhu, Heavy baryons in the relativized quark model with chromodynamics,
\href{https://doi.org/10.1103/PhysRevD.110.056052}{Phys. Rev. D \textbf{110} 5, 056052 (2024)}, arXiv:2405.19039[hep-ph].

\bibitem{P37}G. Yang and J. L. Ping, Dynamical study of $\Omega_{c}^{0}$ in the chiral quark model,
\href{https://doi.org/10.1103/PhysRevD.97.034023}{Phys. Rev. D \textbf{97} 3, 034023 (2018)}, arXiv:1703.08845[hep-ph].

\bibitem{P372}L. Liu, Y. M. Xiao, and T. Guo, Hydrogen-like structures in the strong interaction,
\href{https://doi.org/10.1103/77hs-2gy4}{Phys. Rev. D \textbf{112} 5, 054012 (2025)}, arXiv:2505.22177[hep-ph].


\bibitem{P38}S. Godfrey and N. Isgur, Mesons in a Relativized Quark Model with Chromodynamics,
\href{https://doi.org/10.1103/PhysRevD.32.189}{Phys. Rev. D \textbf{32}, 189-231 (1985)}.

\bibitem{P39}S. Capstick and N. Isgur, Baryons in a relativized quark model with chromodynamics,
\href{https://doi.org/10.1103/PhysRevD.34.2809}{Phys. Rev. D \textbf{34}, 2809-2835 (1986)}.

\bibitem{P40}N.Isgur and G. Karl, P Wave Baryons in the Quark Model,
\href{https://doi.org/10.1103/PhysRevD.18.4187}{Phys. Rev. D \textbf{18}, 4187 (1978)}.

\bibitem{P41}H. H. Zhong, M. S. Liu, and R. H. Ni et al.,Unified study of nucleon and $\Delta$ baryon spectra and their strong decays with chiral dynamics,
\href{https://doi.org/10.1103/PhysRevD.110.116034}{Phys. Rev. D \textbf{110} 11, 116034 (2024)}, arXiv:2409.07998[hep-ph].

\bibitem{P42}G. L. Yu, Z. Y. Li, Z. G. Wang, and Z. Zhou, Systematic analysis of the mass spectra of triply heavy baryons,
\href{https://doi.org/10.1140/epjc/s10052-025-14261-3}{Eur. Phys. J. C \textbf{85}, 543 (2025)},  arXiv:2501.01803[hep-ph].


\bibitem{P43}H. Zhou, S. Q, Luo, and X. Liu, Triply heavy baryon spectroscopy revisited,
\href{https://doi.org/10.1103/jhr1-ccsw}{Phys. Rev. D \textbf{112} 7, 074007 (2025)}, arXiv:2507.10243[hep-ph].

\bibitem{P44}N.Isgur, Meson-like baryons and the spin-orbit puzzle,
\href{https://doi.org/10.1103/PhysRevD.62.014025}{Phys. Rev. D \textbf{62}, 014025 (2000)}, arXiv: hep-ph/9910272.

\bibitem{P45}M. Anselmino, E. Predazzi, and S. Ekelin, et al., Diquarks,
\href{https://doi.org/10.1103/RevModPhys.65.1199}{Rev. Mod. Phys. \textbf{65}, 1199-1234 (1993)}.

\bibitem{P46}W. L. Wu, L. Meng, S. L. Zhu, DeepQuark: A Deep-Neural-Network Approach to Multiquark Bound States
\href{https://doi.org/10.1103/ckpr-s876}{Phys. Rev. Lett. \textbf{136} 7, 071901 (2026)}, arXiv:2506.20555[hep-ph].

\bibitem{P47}E. Hiyama, Y. Kino and M. Kamimura, Gaussian expansion method for few-body systems,
\href{https://doi.org/10.1016/S0146-6410(03)90015-9}{Prog. Part. Nucl. Phys. \textbf{51}, 223-307 (2003)}.

\bibitem{P48}Z. Y. Li, G. L. Yu, Z. G. Wang and J. Z. Gu, Heavy-quark dominance and fine structure of excited heavy baryons $\Sigma_{Q}$, $\Xi'_{Q}$ and $\Omega_{Q}$,
\href{https://doi.org/10.1140/epjc/s10052-024-13706-5}{Eur. Phys. J. C \textbf{84} 12, 1310 (2024)}, arXiv:2405.16162[hep-ph].

\bibitem{P49}G. L. Yu, Z. Y. Li, and Z. G. Wang, et al., Systematic analysis of single heavy baryons $\Lambda_{Q}$, $\Sigma_{Q}$ and $\Omega_{Q}$,
\href{https://doi.org/10.1016/j.nuclphysb.2023.116183}{Nucl. Phys. B \textbf{990}, 116183 (2023)}, arXiv:2206.08128[hep-ph].

\bibitem{P50}Z. Y. Li, G. L. Yu, and Z. G. Wang, et al., Systematic analysis of strange single heavy baryons $\Xi_{c}$ and $\Xi_{b}$,
\href{https://doi.org/10.1088/1674-1137/acd365}{Chin. Phys. C \textbf{47}, 073105 (2023)}, arXiv:2207.04167[hep-ph].

\bibitem{P51}Z. Y. Li, G. L. Yu, Z. G. Wang and J. Z. Gu, Heavy quark dominance in orbital excitation of singly and doubly heavy baryons,
\href{https://doi.org/10.1140/epjc/s10052-024-12457-7}{Eur. Phys. J. C \textbf{84} 2, 106 (2024)}, arXiv:2311.08251[hep-ph].

\bibitem{P52}Z. Y. Li, G. L. Yu, and Z. G. Wang, et al., Mass spectra of singly heavy baryons in the relativized quark model with heavy-quark dominance,
\href{https://doi.org/10.1088/1674-1137/adf178}{Chin. Phys. C \textbf{49} 11, 113107 (2025)}, arXiv:2503.01237[hep-ph].

\bibitem{P53}Z. Y. Li, G. L. Yu, and Z. G. Wang, et al., Mass spectra of doubly heavy baryons in the relativized quark model with heavy-quark dominance,
\href{https://doi.org/10.1140/epjc/s10052-025-15026-8}{Eur. Phys. J. C \textbf{85} 11, 1271 (2025)}, arXiv:2506.19504[hep-ph].




\end{thebibliography}
\end{document}